# The role of the vagus nerve during fetal development and its relationship with the environment

*Francesco Cerritelli[1], Martin G. Frasch[2], Marta C. Antonelli[3,4], Chiara Viglione[1], Stefano Vecchi[1], Marco Chiera[1*] and Andrea Manzotti[1,5,6]*

Affiliations

[1] RAISE lab, Foundation COME Collaboration, 65121 Pescara, Italy

[2] Department of Obstetrics and Gynecology and Center on Human Development and Disability(CHDD), University of Washington, Seattle, WA, USA

[3] Instituto de Biología Celular y Neurociencia "Prof. E. De Robertis". Facultad de Medicina, UBA, Buenos Aires, Argentina.

[4] Department of Obstetrics and Gynecology, Klinikum rechts der Isar, Technical University of Munich, Germany.

[5] Department of Pediatrics, Division of Neonatology, "V. Buzzi" Children's Hospital, ASST-FBF-Sacco, 20154 Milan, Italy.

[6] Research Department, SOMA, Istituto Osteopatia Milano, 20126 Milan, Italy

**\*Correspondence:**
Marco Chiera
marco.chiera.90@gmail.com

Word count: 16,009
Tables: 1
Figures: 0

# Abstract

The autonomic nervous system (ANS) is one of the main biological systems that regulates the body's physiology. ANS regulatory capacity begins before birth as the sympathetic and parasympathetic activity contributes significantly to the fetus' development. In particular, several studies have shown how vagus nerve is involved in many vital processes during fetal, perinatal and postnatal life: from the regulation of inflammation through the anti-inflammatory cholinergic pathway, which may affect the functioning of each organ, to the production of hormones involved in bioenergetic metabolism. In addition, the vagus nerve has been recognized as the primary afferent pathway capable of transmitting information to the brain from every organ of the body. Therefore, this hypothesis paper aims to review the development of ANS during fetal and perinatal life, focusing particularly on the vagus nerve, to identify possible "critical windows" that could impact its maturation. These "critical windows" could help clinicians know when to monitor fetuses to effectively assess the developmental status of both ANS and specifically the vagus nerve. In addition, this paper will focus on which factors – i.e. fetal characteristics and behaviors, maternal lifestyle and pathologies, placental health and dysfunction, labor, incubator conditions, and drug exposure – may have an impact on the development of the vagus during the above-mentioned "critical window" and how. This analysis could help clinicians and stakeholders define precise guidelines for improving the management of fetuses and newborns, particularly to reduce the potential adverse environmental impacts on ANS development that may lead to persistent long-term consequences. Since the development of ANS and the vagus influence have been shown to be reflected in cardiac variability, this paper will rely in particular on studies using fetal heart rate variability (fHRV) to monitor the continued growth and health of both animal and human fetuses. In fact, fHRV is a non-invasive marker whose changes have been associated with ANS development, vagal modulation, systemic and neurological inflammatory reactions, and even fetal distress during labor.

# Table of content

# 1. Introduction

The autonomic nervous system (ANS) is one of the main biological systems that regulates the body's physiology: in particular, the interaction between its two branches – the sympathetic nervous system (SNS) and the parasympathetic nervous system (PNS) – allows the organism to efficiently cope with endogenous and exogenous stressors (Goldstein, 2006; Bonaz et al., 2021).

Regarding the PNS, the vagus nerve has been recognized as a fundamental afferent pathway for sensing the milieu interior in peripheral tissues, thus allowing the organism to respond by activating complex networks involving the whole ANS, endocrine, metabolic, innate and adaptive immune systems (Frasch et al., 2018b; Bonaz et al., 2021).

One of such responses, able to perform homeokinetic adjustments, is the cholinergic anti-inflammatory pathway (CAP). Briefly, through the vagus nerve, the brain receives information about the body's inflammatory milieu and, thus, activates neuroimmune responses to dampen the detected inflammation (Garzoni et al., 2013; Frasch, 2020; Bonaz et al., 2021). Very little is known, however, about the central representation, the information processing network, of the vagus nerve. An idea of neuroimmunological homunculus has been proposed and requires further research (Conway et al., 2006; Tracey, 2007; Diamond and Tracey, 2011; Frasch et al., 2018a; Castel et al., 2020).

The ANS regulatory capacity begins before birth as the sympathetic and parasympathetic activity contributes significantly to the fetus' development (Hoyer et al., 2017; Mulkey and du Plessis, 2018; Zizzo et al., 2020). Indeed, several studies have shown the vagus nerve is involved in many vital processes during fetal, perinatal and postnatal life: from the regulation of inflammation through the CAP, which may affect each organ (Herry et al., 2019; Frasch, 2020), to the production of hormones involved in bioenergetic metabolism (Bystrova, 2009).

On the other hand, the ANS activity, as measured through heart rate variability (HRV) – a non-invasive biomarker evaluating the variability of the time intervals between successive heartbeats – is affected by the developmental and health conditions of both fetus and infant. HRV analysis can give paramount hints about the fetus' or newborn's wellbeing and allow clinicians to predict the occurrence of even deadly complications (Table 1) (Stone et al., 2013; Al-Shargabi et al., 2017; Massaro et al., 2017; Chiera et al., 2020; Frasch, 2020).

Nonetheless, the importance of ANS and vagus nerve regulation in fetal and neonatal development is still underrepresented (Sadler, 2012; Moore et al., 2016). Moreover, except the third trimester (Mulkey and du Plessis, 2018, 2019), literature usually lacks an analysis of potential "critical windows" in the perinatal period that may give information about ANS development. Defining such critical windows during gestation could help clinicians know exactly when to monitor fetuses to effectively assess the developmental status of both ANS and vagus nerve.

Therefore, the present paper aims to review the fetal and perinatal development by focusing on the development of the ANS and, in particular, of the vagus nerve to identify possible critical windows that could impact their maturation. This paper will also focus on which factors – i.e. fetal characteristics and behaviors, maternal lifestyle and pathologies, placental health and dysfunction, labor, incubator conditions, and drug exposure – may affect the development of the vagus.

## 2. The fetal ANS development: the search of "critical windows"

As the ANS is controlled by several brain areas (e.g., amygdala, hypothalamus, insula, cingulate cortex, and several brainstem nuclei) that constitute the central autonomic network (CAN), paying attention to their development is fundamental for studying the fetal ANS. In the same way, due to the CAP, it is also relevant to review the development of the organs potentially connected to the vagus nerve (Sklerov et al., 2019; Frasch, 2020). As the vagus nerve affects the body metabolism in a pleiotropic manner, such a review could help better understand the networks underlying this effect and how to harness them for improving fetal and neonatal health (Castel et al., 2020).

### 2.1. A review of ANS development

#### 2.1.1. The development during the 1st trimester

The nervous system differentiates from the ectoderm approximately around 3 weeks gestational age (WGA), when part of the ectoderm develops into the neural plate. At 4 WGA, neurulation occurs: the neural folds arise from the neural plate and begin to fuse in both cranial and caudal direction, thus forming the neural tube, that is, the future central nervous system (CNS). During neurulation, some neuroectodermal cells detach from the neural folds and differentiate into the neural crests (NCs), the specialized cells that will form the future ANS (Moore et al., 2016).

By 4 WGA, the forebrain, midbrain, and hindbrain (the primary brain vesicles) are already visible, as are the spinal ganglia that come from the NCs. In the next weeks, the nervous system rapidly grows: the five secondary brain vesicles, along with the four ventricles and the meninges form, while the glioblasts (the glial cells precursors) start to differentiate and migrate throughout the developing nervous system. The neural tube also begins to thin to form the central canal of the spinal cord (Moore et al., 2016).

During this period (4-8 WGA), from the NCs several derivatives arise: the dorsal root ganglia, the sympathetic ganglia, the renal, celiac, and intestinal ganglia and plexus, the afferent ganglia of cranial nerves, and the suprarenal medulla (Newbern, 2015; Moore et al., 2016). In particular, by 6 WGA, the first parasympathetic ganglia can be detected (Müller and O'Rahilly, 1989) and, by 10 WGA, the sympathetic trunks have appeared throughout the vertebral column (Kruepunga et al., 2021) and, apart from the colonic loop, the extrinsic innervation of the enteric nervous system (ENS) is complete (Kruepunga et al., 2020). NC cells also contribute to spleen development (Barlow-Anacker et al., 2017). The spleen begins to develop at 5 WGA as a lobulated organ, appears clearly around 8 WGA, and initially functions as a hematopoietic center; then, during fetal development, it gradually loses both this role and its lobules (Moore et al., 2016; Hill, 2021a).

Among all the neural crest cells, the vagal NC cells are of particular interest due to their derivatives. Vagal NC cells emerge adjacent to the somites 1-7, rostral to the head NCs and caudal to the trunk and sacral ones, and are the ENS precursors (Hutchins et al., 2018; Simkin et al., 2019). But these cells also contribute to the formation of organs such as the heart (especially, the intrinsic cardiac nervous system), thyroid and parathyroids, thymus, lung, and pancreas ganglia. Vagal NC cells then give rise to the parasympathetic, sympathetic and sensory ganglia (Hutchins et al., 2018; Simkin et al., 2019; Fedele and Brand, 2020).

Several animal studies in vertebrates and mammals show that vagal NC cells ablation heavily impairs the development and function of the aforementioned tissues. For example, neurons and glial cells in the pancreas do not appear and the thymus loses both its vasculature and its capacity of developing T-cells (Hutchins et al., 2018).

At 5-6 WGA, the twelve cranial nerves appear (Moore et al., 2016). As the parasympathetic system is constituted by the oculomotor, facial, glossopharyngeal, and vagus nerves, and since the trigeminal, facial, glossopharyngeal, spinal accessory, and vagus nerves lay the foundations for social interaction especially in newborns – these nerves are central to sensing environmental stimuli, including the human voice, and to carrying out behaviors such as head-turning, smiling, sucking and swallowing – (Porges, 2011), the cranial nerves maturation may be paramount for an optimal ANS growth and, thus, for developing regulatory adaptive responses. Regarding the vagus nerve, at this stage, the dorsal motor nucleus is the predominantly active effector center (Porges, 2011).

At 4-6 WGA the primitive diaphragm appears (Hill, 2021d) and, by 7 WGA, the four chambers of the heart have formed, thus connecting the heart to the aorta and the pulmonary trunk. By 10 WGA, fetal circulation passively exchanges gas mainly with the placenta and breathing begins. As gestation proceeds, blood entering the pulmonary circulation increases (Morton and Brodsky, 2016; Tan and Lewandowski, 2020).

From 7-8 WGA gray and white matter differentiate and, by the end of the 1st trimester, after the embryo has become a fetus, it is possible to recognize all the main brain structures, including the choroid plexus, thalamus, hypothalamus, amygdala, basal ganglia, corpus callosum, brainstem nuclei, cerebellum, and the neural plate of the insula (Moore et al., 2016; Hill, 2021b). Around the transition between the 1st and 2nd trimester, even the circumventricular organs, in particular, the area postrema, appear: they are organs devoid of the brain-blood barrier that, hence, allow the brain to be directed informed about the biochemical status of the internal milieu (Gokozan et al., 2016).

Interestingly, at 6 WGA, the embryo begins to show spontaneous movements and can "respond" to touch and light, as the motor spinal nerve fibers have appeared around 4 WGA (Moore et al., 2016): indeed, at 8 WGA the embryo is capable of activating a spinal reflex in response to touch, and between 7-11 WGA all facial reflexes have developed (Borsani et al., 2019). At 10 WGA the corticospinal connectivity begins to develop, together with an increase in synaptogenesis, and, as a result, body reflexes in the upper limbs start to appear (Borsani et al., 2019). Since fetal movements (FMs) tend to synchronize with fetal HRV (fHRV), their detection can give important insights on ANS development, even in the 1st trimester (DiPietro et al., 2007; Zizzo et al., 2020). From 12-13 WGA, fHRV-FMs coupling becomes heavily affected by external factors: therefore, its analysis could shed more light on how the fetus is affected by maternal and environmental factors (DiPietro et al., 2015).

During this period, yawning develops (Borsani et al., 2019). Since yawning has been related to thermoregulatory behaviors and the ability of "switching" between arousal and rest states, both of which are strictly dependent on ANS functionality, the detection of yawning at this early age could represent a useful indicator of ANS development or, as suggested, brainstem functionality (Walusinski, 2006, 2010; Gallup and Eldakar, 2013; Walusinski, 2014). This hypothesis could be reinforced by the evidence that yawning seems to rely on interoception, the inner sense that constantly monitors the internal milieu and for which the vagus nerve plays a central role, and to be processed by the CAN (in particular, insula, hypothalamus, locus coeruleus, and reticular activating system) (Walusinski, 2006). Lastly, yawning amount and duration seem to correlate with neuronal development in mammals (Gallup et al., 2016).

### 2.1.2. The development during the 2nd trimester

At the beginning of the 2nd trimester, the cerebral cortex creates more and more sulci and gyri to increase its surface area and allow brain development. In particular, the subplate (a central hub for cortical connectivity) becomes visible and begins to receive connection from the thalamus, basal forebrain, and brainstem. Around 17-19 WGA, the cerebellum development can be assessed through magnetic resonance imaging (Hill, 2021d), the corticospinal tracts towards the whole body appear, and at 20 WGA the spinothalamic tract is completed (Borsani et al., 2019).

At 18-19 WGA, nociceptive neurons and the typical hormonal response to pain appear (Borsani et al., 2019; Chen et al., 2021), thus hinting that primitive pain perception might begin at this stage. Although thalamocortical connections, which strongly increase around 20-22 WGA, are usually required for pain experience and the hemodynamic-behavioral responses to nociceptive stimuli are usually detected around 26 WGA (Verriotis et al., 2016; Chen et al., 2021), several authors argue that fetuses could feel pain at least from 12 WGA, as the first superficial thalamic-subplate connections seem to be functionally equivalent to the later thalamocortical ones (Derbyshire and Bockmann, 2020).

Around 20 WGA and for the rest of the 2nd trimester, myelination occurs (Borsani et al., 2019). From the neural crests, neurilemma cells arise and form the myelin sheaths that wrap both the axons of somatic motor neurons and the preganglionic autonomic motor neurons (Moore et al., 2016). Also, fetal microglia from the bone marrow invade the CNS, while gray and white brain matters keep on developing: the cerebral lobes with all their sulci and gyri, cortexes such as the insula, the cerebellum, the pons, and many other structures increasingly resemble their adult counterparts (Moore et al., 2016). From 17 to 24 WGA, the spleen almost completely develops its reticular framework (Satoh et al., 2009; Hill, 2021a). Regarding the ANS, brainstem and hypothalamic centers form several connections with the other CAN structures, that is, insula, anterior cingulate gyrus, hippocampus, hypothalamus, and amygdala (Mulkey and du Plessis, 2018; Schlatterer et al., 2021).

During the 2nd trimester, the body becomes covered with lanugo, a fine downy hair that plays a paramount role in neurodevelopment. Indeed, through lanugo, the flow of amniotic fluid and FMs stimulate the low-threshold unmyelinated C afferent fibers (Bystrova, 2009), which have been found to be essential for both nociception and interoception . This stimulation reaches the brainstem, in particular the nucleus of the solitary tract – the main afferent nucleus of the vagus nerve – but also other nuclei including the parabrachial ones, and then more cortical structures such as the hypothalamic paraventricular nucleus and the posterior insular cortex. As a result, several motor and neuroendocrine responses are induced: for instance, gastrointestinal and anabolic hormones such as cholecystokinin, insulin, and insulin-like growth factor-1 are released. On the other hand, the hypothalamic-pituitary-adrenals (HPA) axis and vagal activation become more finely tuned (Bystrova, 2009).

Regarding sensorimotor development, around 15 WGA a complex set of movements (e.g., breathing movements, general body movements, isolated limb, head, and neck movements, startle and twitch movements, and swallowing) are available, whereas, by 18

WGA, stable handedness can be detected. Then, as gestation progresses, the amount of movement, both spontaneous and in response to vibroacoustic stimuli, gradually increases: indeed, mothers begin to feel their babies' movements (Moore et al., 2016; Borsani et al., 2019). At about 18-20 WGA, rapid eye movements (REM) can be detected (Borsani et al., 2019) as well as quiet and active sleeping states (Dereymaeker et al., 2017), although fetal behavioral states do not show particular coherence with ANS tone (Brändle et al., 2015).

As several vital systems are forming around the end of the 2nd trimester (e.g., alveolar ductal development and surfactant secretion begin at 24 WGA), babies born prematurely before 26 WGA have a higher chance of dying or developing neurodevelopmental disability (Moore et al., 2016; Morton and Brodsky, 2016; Hill, 2021c). Nevertheless, neonatal intensive care units have improved so much in the years that babies born as early as 20-22 weeks can survive if properly treated (Moore et al., 2016).

It is worth noting that, from the 2nd trimester, HRV metrics can successfully estimate fetal autonomic brain age (Hoyer et al., 2012, 2013b, 2015, 2017, 2019). Interestingly, gestational age can also be precisely estimated by measuring the opening and closing time of fetal cardiac valves (Marzbanrad et al., 2017), finding that could highlight the contribution of intrinsic fHRV or, at least, heart activity, to global HRV.

### 2.1.3. The development during the 3rd trimester

At 26-29 WGA, fetuses become able to efficiently control body temperature (Moore et al., 2016). By 29 WGA, all the major corticospinal, spinothalamic, and spinocortical tracts as well their myelination are completed, whereas from 24 WGA functional connectivity patterns in the fetal brain are recognisable. Indeed, at 32 WGA the cerebral cortex resembles the adult one and, in the next week, synchronicity between the hemispheres develops, as well as complex connections among several brain areas and networks (Moore et al., 2016; Borsani et al., 2019). As the 3rd trimester advances, specific nociceptive event-related potentials in the brain increasingly develop until 35-37 WGA (Verriotis et al., 2016).

From 25 WGA until birth, the vagus nerve myelination, the vagal control by the nucleus ambiguus, and, thus, the PNS rapidly develop, as testified by the appearance of short-term variability in time-domain HRV metrics (e.g., HF) and the proportional reduction of long-term variability metrics (e.g., LF), more related to SNS and baroreflex functions (Mulkey and du Plessis, 2018, 2019; Schlatterer et al., 2021). In particular, a steep rise in PNS tone can be detected around 37-38 WGA and basal fetal cardiovascular function becomes more influenced by the PNS. Nevertheless, from 30 WGA, cortisol, thyroxine, and catecholamines levels rise to prepare the fetus for birth. In the case of preterm birth, the ANS remains predominantly affected by both the SNS and the primitive vagal dorsal motor nucleus, with consequences on the stress response regulation towards environmental stimuli (e.g., excessive tachycardia or bradycardia) (Morton and Brodsky, 2016; Mulkey et al., 2018; Mulkey and du Plessis, 2019).

Lastly, whereas general movements (GMs) tend to increase until 32 WGA and then decrease as birth approaches, facial movements increase more and more (Borsani et al., 2019). During the 3rd trimester, distinct fetal behavioral states (i.e., quiet sleep, active sleep, quiet awake, and active awake) become increasingly recognisable through the analysis of heart rate, eye movement, and GMs, as corticothalamic and brainstem areas develop more complex interactions (Brändle et al., 2015; Borsani et al., 2019). Interestingly, the analysis of specific HRV metrics, for instance, SDNN (more related to global neuroautonomic regulation), RMSSD (more related to vagal modulation), and PE, can help identify the actual behavioral state and, therefore, detect potential problems in neurodevelopment whether incoherence between HRV and body movements is detected (Schneider et al., 2008; Brändle et al., 2015).

### 2.1.4. The development during labor

Brith, as the transition from fetal to neonatal physiology, is a major challenge for the fetus and requires an ANS able to efficiently adapt the cardiovascular, respiratory, thermoregulatory, and metabolic (e.g., glycemic regulation) systems to the new environment. For this reason, before birth neuroendocrine signalling (e.g., catecholamines, cortisol, thyroid hormones, and renin-angiotensin) surges and SNS outflow from CAN (e.g., hypothalamus, cortical, and brainstem centers) greatly increases. During labor, however, the nervous system is highly vulnerable to injuries due to hypoxia, energy deprivation, or oxidative stress (Morton and Brodsky, 2016; Mulkey and du Plessis, 2018; Mulkey et al., 2021).

Compared to cesarean section (CS), spontaneous labor without analgesia seems to improve cardiovagal modulation as shown by HF increase and LF decrease, despite several HRV metrics tend to increase in the postnatal days regardless of delivery mode (Kozar et al., 2018). A well-developed ANS responds efficiently to umbilical cord clamping by inducing the fetus to commence spontaneous breathing. Otherwise, especially in case of a difficult delivery or CS without labor, the fetus may experience hypoxia and have difficulties in secreting pulmonary surfactant, thus developing respiratory pathologies that can backfire, in particular, on ANS brainstem centers through oxygen deprivation and hinder autonomic regulation (Hillman et al., 2012; Morton and Brodsky, 2016; Mulkey and du Plessis, 2018).

It is noteworthy that all the adaptations required to pass from intrauterine to extrauterine life seem to rely on cortisol, whose release is markedly reduced in case of preterm birth or lack of labor (Hillman et al., 2012; Morton and Brodsky, 2016). Since cortisol is an efferent agent in anti-inflammatory neural reflexes like the CAP (Bonaz et al., 2021), its reduced production may alter the CAP development itself. Indeed, low cortisol levels could fail in regulating postnatal inflammation, e.g., necrotizing enterocolitis (NEC), which could induce cell death in the vagal brainstem centers (Fritze et al., 2014).

### 2.1.5. The development in the postnatal life

After labor, pulmonary vascular resistance decreases compared to systemic vascular resistance as lungs inflate and fetal circulation becomes active (Tan and Lewandowski, 2020), whereas the nervous system integrates sensory, motor, and regulatory functions in more complex behaviors (Moore et al., 2016). In particular, centers that are paramount to neurotransmitters (e.g. noradrenaline) biosynthesis and autonomic functions (e.g., temperature, breathing, and satiety regulation) including the area postrema greatly develop around birth (Gokozan et al., 2016).

As revealed by HRV analysis, the ANS constantly develops. Indeed, several metrics related to both SNS and PNS (e.g., SDNN, CSI, HFn, LFn, TP, SD2, HRV Index, and Parseval Index) change right after birth and in the following months, with the PNS modulation that takes predominance (e.g., lower LF/HF and higher RMSSD and HF) during the following 2 years (Oliveira et al., 2019; Patural et al., 2019) to better regulate the increasingly social interactions the infant encounters (Porges, 2011).

It is worth remembering that the two ANS branches are anything but antagonists: in fact, the regulation of many organic functions (Goldstein, 2006), including the immune and inflammatory responses (Bonaz et al., 2021), strictly depends on the synergy between the two branches. Should one of them be damaged, the other one alone could not regulate the body's homeostasis efficiently anymore.

At birth, the higher cortical networks controlling the ANS increase their connectivity (Schlatterer et al., 2021) and, in low-risk newborns at around 1 day of age, coherence between electrocortical activity and brainstem-mediated autonomic tone, especially parasympathetic, can be easily detected (Mulkey et al., 2021). Social interactions, including skin-to-skin contact between mother and newborns, help increase ANS maturation as well as stress responsivity and circadian rhythms, in particular in preterm infants (Ulmer Yaniv et al., 2021).

As newborns spend most of their time sleeping, ANS development correlates also with sleep development, with quiet sleep (absence of movements and slow rhythmic breathing) characterized by PNS predominance and active sleep (myoclonic twitching, facial, eye, and head movements, and irregular heart and breath rates) characterized by SNS predominance. Hence, measuring HRV during different sleep states can help better assess ANS development (Yiallourou et al., 2012; Dereymaeker et al., 2017).

## 2.2. How disruptions in "critical windows" can affect ANS development

In the previous section, we described fetal development with particular attention to ANS, CAN, sensorimotor development, and some organs. We have shown how alterations in CAN activity and sensorimotor behaviors can give hints about the developing ANS function. Similar hints can also be gained by monitoring organ development due to the research regarding the CAP and the systemic interactions of the vagus nerve (Bonaz et al., 2021).

However, the intrinsic spatiotemporal complexity of ANS development and the complex species differences in the timelines of neural development from which inferences are often drawn to human development make it difficult to identify specific or exclusive critical windows, although we might point to particular periods: 4-8 WGA, due to organogenesis; 12 WGA, due to subplate and fHRV-FMs coupling appearance; 18-24 WGA, due to nociception, CAN, and spleen development and myelination; the 3rd trimester, due to PNS development and change in FMs and hormones production; and birth, due to ANS adaptation to interact directly with the external environment.

Therefore, the next two sections will discuss short-term and long-term consequences of disruptions in ANS development.

### 2.1. Short-term consequences

Fetal or birth complications affecting ANS development may have consequences on organic functions and mortality right after birth and in the following months.

ANS impairment occurring in utero due to congenital heart disease, chronic hypoxemia, fetal growth restriction, acidemia, the trial of labor, bacterial and viral infections or maternal factors can lead to labor complications (even death) and prematurity (Mulkey and du Plessis, 2018). Since the ANS requires 37 weeks in utero to fully develop, preterm newborns can experience difficulties in regulating their internal milieu in the face of environmental factors (Praud et al., 2017). Moreover, preterm birth often involves invasive treatments (e.g., invasive respiratory support) or drugs that can stress even more the immature ANS. Indeed, ANS responsiveness seems to remain quite low even when preterm infants reach term-equivalent ages compared to full-term newborns (Mulkey et al., 2018; Patural et al., 2019; Schlatterer et al., 2021).

Health conditions may, however, affect postnatal ANS maturation more than gestational age at birth: in the three months after birth, infants with a higher number of postnatal complications (e.g., infections, patent ductus arteriosus requiring treatment, mechanical ventilation) show a lower ANS tone measured through DFA metrics, i.e., α1, α2, RMS1, and RMS2. Whereas α1 differs already at birth, the other metrics show different trajectories after 30-40 days, hinting at different ANS development due to the experienced postnatal comorbidities (Schlatterer et al., 2021). Asymptomatic toddlers born to mothers infected with Zika virus during pregnancy, show alterations in their nonlinear HRV measures reflecting chaotic behavior and recurrence plot properties. These changes were suggested to result from the effects of chronic hypoxia – which Zika virus infection is known to cause – on the interactions between the ANS and the cardiac pacemaker cell development which generates intrinsic HRV (Frasch et al., 2020a; Herry et al., 2021).

It is widely known that comorbidities such as NEC or hypoxic-ischemic encephalopathy (HIE) can negatively affect ANS (Schlatterer et al., 2021). For instance, stroke can easily alter both sympathetic and parasympathetic tone, thus impairing the ANS regulatory capacities (Reich et al., 2019). Specific kinds of brain injuries seem also to specifically affect ANS: for instance, DFA revealed that white matter injury, watershed stroke, basal ganglia injury, or global injury can show characteristic HRV signatures (Metzler et al., 2017). Prematurity increases the risk of these adverse events (Mulkey and du Plessis, 2018). Fetal gut injury due to inflammation in a fetal sheep model of NEC etiology leaves a specific HRV signature and preterm neonates show abnormalities in HRV 1-3 days before clinical symptoms (Stone et al., 2013; Liu et al., 2016).

However, we have yet to fully understand ANS development, even after birth. Despite knowing that the CAN coordinates autonomic function and that left and right hemispheres modulate PNS and SNS tone respectively (Mulkey et al., 2021), we have conflicting results on how the brain precisely controls ANS in newborns. For instance, whereas hypoxic-ischemic injuries in the right hemisphere seem to depress SNS tone, vaso-occlusive strokes in the same hemisphere appear to increase SNS tone. Whereas vaso-occlusive strokes in the left hemisphere seem to increase PNS tone, hypoxic-ischemic injuries seem to have no effect on PNS tone. In some cases, insular injuries do not significantly change ANS tone; or, despite the effects on ANS tone, systemic hemodynamic meters (e.g., diastolic and systolic blood pressure) do not seem affected by brain injuries (Schneebaum Sender et al., 2017; Reich et al., 2019).

Many postnatal pathologies – e.g., hyperbilirubinemia, congenital heart disease, respiratory distress syndrome, respiratory syncytial virus, NEC, and sepsis – can lead to ANS impairment, probably due to inflammatory cytokines and toxins that can bypass the brain-blood barrier and induce neural death in areas paramount to ANS control (Fritze et al., 2014; Al-Shargabi et al., 2017; Bonaz et al., 2021).

As pathologies can affect ANS development, altered ANS function can make newborns more vulnerable to those same pathologies, due to failure in activating efficient anti-inflammatory reflexes including the CAP (Yiallourou et al., 2012; Mulkey et al., 2018; Bonaz et al., 2021). Indeed, dysfunctional CAP can fail in regulating inflammation, thus letting the immune system go unchecked and free to induce tissue damage, shock, and even death, whereas, on the other hand, a functional CAP can prevent microglia activation, thus showing a neuroprotective effect (Garzoni et al., 2013; Frasch et al., 2016).

ANS impairment, maybe through impaired cardiocirculatory and ventilatory regulation, can increase the risk of apparent life-threatening events or sudden infant death syndrome (SIDS) and, thus, of death in the first 6-12 months of postnatal life. Indeed, newborns that died from SIDS seemed to have brainstem and cerebellar impairments and to fail in arousing from sleep, which could highlight the importance of assessing HRV development during pregnancy (Yiallourou et al., 2012; Horne, 2018; Mulkey et al., 2018).

It is interesting that fetal ANS impairment can also have consequences on infant temperament at both 3 and 6 months (Howland et al., 2020).

## 2.2. Long-term consequences

ANS impairment may have consequences on the global development that persist during infancy, adolescence and adult life.

Although preterm infants may reach good ANS maturation after 2-3 years (De Rogalski Landrot et al., 2007), ANS impairment tends to remain even later in life and to be accompanied by SNS and HPA regulation impairment, both during wakefulness and sleep (Haraldsdottir et al., 2018; Urfer-Maurer et al., 2018).

ANS dysfunction in the first year of life could have consequences on cardiovascular regulation even during adolescence, which makes paramount monitoring ANS development in the 1st and 2nd year to discover infants at risk. Indeed, cardiovascular regulation impairment could lead to higher mortality due to cardiovascular events (Patural et al., 2019). An immature ANS due to inflammation, prematurity, or fetal growth restriction can also lead to other pathologies such as type 2 diabetes (Mulkey and du Plessis, 2018). These and other metabolic or immune pathologies that could arise later in life might have their origin in alteration of CAP functioning, that is, in inflammation regulation by the ANS (Pavlov and Tracey, 2012; Burgio, 2015; Bonaz et al., 2021). Furthermore, ANS alterations in newborns can lead to neurological dysfunctions including cerebral palsy at 2 years of age (Dimitrijević et al., 2016).

Since ANS development is coupled with FMs development, ANS alterations can heavily affect CNS and neurobehavioral growth: on the one hand, FMs are usually considered the base for the fight-or-flight response, thus influencing the stress response (Schmidt et al., 2014); on the other hand, ANS immaturity correlates with altered cerebral and psychomotor development (Hoyer et al., 2014; Schmidt et al., 2014; Schneider et al., 2018), and even with altered cognitive, language, playing, and social skills (Doussard-Roosevelt et al., 1997, 2001; Bornstein et al., 2002; DiPietro et al., 2007; Siddiqui et al., 2017).

ANS immaturity due to fetal or postnatal complications (e.g., maternal pathologies, birth difficulties) could then impair the whole brain development and, therefore, behavior, stress response, and mood regulation, with negative consequences – even serious neurological or psychological pathologies – in infants, adolescent, and adult life (Mulkey et al., 2018; Mulkey and du Plessis, 2019). Even painful procedures during hospitalization seem to induce alteration in stress regulation and brain development in newborns (Holsti et al., 2006). Maternal stress during pregnancy alters fetal ANS by entraining fetal heart rate with maternal heart rate decelerations during respiratory effort (Lobmaier et al., 2020). 8-years old children with autism spectrum disorders (ASD), depression or conduct disorder showed respectively distinct signatures of HRV compared to controls and it has been proposed that, at least for ASD etiology, such abnormalities in HRV may be detected as early as in utero where they may be induced by asymptomatic inflammatory events altering ANS, among other systems (Frasch et al., 2019, 2021).

Due to the negative consequences ANS impairment can have later in life, it is paramount to rely on any protective factor that can sustain ANS maturation, starting from taking care of maternal health and lifestyle, which can greatly affect fetal development and, then, breastfeeding. In the same way, stress relief during pregnancy, skin-to-skin contact and social interactions between newborns and parents/caregivers can help newborns tune their ANS, reduce stress, increase resilience and improve their neural plasticity (Horne, 2018; Mulkey and du Plessis, 2018; Manzotti et al., 2019; Antonelli et al., 2021).

# 3. The environmental factors affecting the ANS development

Whereas in the previous section we described the fetal development in relation to ANS maturation, showing potential short-term and long-term alteration in case something should negatively affect pregnancy or labor, this section will now detail what factors can alter ANS development and in which ways.

## 3.1. Prenatal life influence on the ANS

### 3.1.1. Fetal factors

#### 3.1.1.1. Incorrect fetal movements

Around 8-9 WGA, it is possible to clearly record GMs (Prechtl 1990). GMs analysis can differentiate specific movement patterns, concerning a specific part of the fetal body, from non-specific movement patterns, involving more than one body segment (de Vries et al., 1982, 1985). GMs quality, as well as their variability and modulation in intensity and velocity, could be valid indicators of fetal neurodevelopment, in particular concerning supra-spinal control on motor activity (Moreira Neto and Porovic, 2018). Different methods could be used to evaluate the quantity and quality of FMs, such as mother perception, acceleration and variability at cardiotocography (CTG), real-time ultrasound, myography, actocardiography (Zizzo et al., 2020), and the Non-stress Test, which relates FMs to fetal heart rate (fHR) accelerations (Bryant et al., 2020).

Both gross motor activity and breathing FMs, particularly during early gestation, represent an important evaluative parameter for fetal neurobehavior and to predict future neurodevelopmental disorders. Indeed, respiratory sinus arrhythmia appears to be strictly related and modulated by efferent vagal activity. Fetal motor activity, therefore, results closely tied to breathing and cardiovascular patterns, which are modulated by the ANS (Zizzo et al., 2020).

According to a recent systematic analysis, Fetal movements can affect the ANS as revealed through modifications in the time-domain and frequency-domain fHRV metrics. The onset of fetal breathing movements, in particular, induces a variation in the main vagal nerve activity parameters: RMSSD and HF power increase, while LH/HF power decreases. Variations are also observed for SDNN, SDNN/RMSSD, LFn, and HFn (Zizzo et al., 2020). In this sense, monitoring the quality and the evolutionary pattern of FMs could reveal useful insight regarding ANS, in particular, vagal maturation.

FMs patterns could be limited as a consequence of an inappropriate ratio between fetal and uterine dimension (Shea et al., 2015; Verbruggen et al., 2018), as well as a consequence of an inadequate amount of amniotic fluid (Sival et al., 1990). Since we have seen that amniotic flow and fetal growth restriction can affect vagal development, their effects could then be mediated by FMs. Indeed, fetuses with intrauterine growth restriction (IUGR) frequently present an oxygenation impairment, with a decrease of breathing movements and GMs, and an increasing number of heart rate (HR) decelerations (Bekedam et al., 1991).

Fetal motor capacities, through the implied neuronal activity, is paramount for differentiation, migration, synaptogenesis, and development of neuronal networks.

Environmental stimuli, per se, induce motor reactions, as the sensory system maturation proceeds. GMs allow fetuses to experience their bodies and the surrounding space, hence exploring the consequences their movements have on their bodies and environment (Fagard et al., 2018). In summary, the FMs make it possible to start developing a primitive body map. The repetitive movements, with the consequent repetitive sensory afferents induced, produce habituation in the fetus, that is manifested also in fHR changes: the ANS trains to adapt the body systems' response to environmental stimuli, even through the movements (Lecanuet et al., 1992; Kisilevsky and Low, 1998).

It is therefore legitimate to argue that a decreased possibility/capacity to express movements by the fetus, that is related to an impairment of the sensory stimulation, could interfere with the normal neurodevelopmental pattern of ANS and vagus nerve itself.

Maternal perception of FMs, in conclusion, is one of the factors that could influence in a relevant way the emotional state of the mother. Despite the fact that there are contrasting results about the sensitivity of fetal activity reported by the mother (Sorokin et al., 1982; Schmidt et al., 1984), it appears that there is an association between the increase of force and frequency of perceived movements and a reduced risk of stillbirth and negative outcomes (Heazell et al., 2018). In some cases, the healthcare providers recommend around 28 WGA the "count to 10" method: the mother counts the number of fetal movements every day, at the same hour, verifying that they occur not less than 10 times in a 2–3-hour period (Bryant et al., 2020). Moreover, it seems that monitoring fetal movements could contribute to creating a primitive maternal-child bonding, even before birth (Flenady et al., 2019). Diminished perception of this motor activity, could therefore be worrisome for the mother, leading to an increased amount of stress, which in turn may have an impact on fetal neurodevelopment (Bryant et al., 2020).

### 3.1.1.2. Fetuses' sex

A variety of different ANS and vagus nerve pattern activities are now well established between different sexes, which can be well represented by HRV tracking even during fetal life. Regarding vagal activity in fetuses, not all the authors agree in observing significative sex-related differences (Oguch and Steer, 1998; DiPietro et al., 2004; Lange et al., 2005; Bhide and Acharya, 2018), even when the same methodologies were used. The discrepancy between these studies may be partly due to the different experimental variables.

Male fetuses show a more reactive ANS and less complexity in control systems than female ones, with a more linear and significantly less complex fHR activity (Bernardes et al. (Bernardes et al., 2008). Moreover, an exploratory study conducted on twins, highlighted that sex-specific differences in fetuses can be noticed both in linear and non-linear fHR metrics. This seems to be significantly influenced also by the combination between twins of different sexes, both in intrapair average and absolute differences (Tendais et al., 2015).

The regulatory patterns of sympathovagal balance show sex-specific differences as pregnancy progresses (Buss et al., 2009). During the 1st trimester, there are no significant differences in HRV track between male and female fetuses (McKenna et al., 2006). Both sexes demonstrate an evolutionary pattern with increased entropy indexes until 34 WGA, with a slight decrease after 37 WGA. With the passing of the weeks, however, sex-specific differences in the sympathovagal balance start to be detected. From 34th WGA, female fetuses show indeed a higher mean fHR and entropy, as well as a lower short-term variability and sympathovagal balance, compared to the males (Gonçalves et al., 2017). This could suggest a different developmental pattern of the vagus nerve, both sex- and time-specific.

In this sense, hormones may play a crucial etiological role, as literature shows that gonadal hormones have a significant impact on the sex-dependent differentiation of the entire nervous system (Chung and Auger, 2013). During pregnancy, the maturation pattern of fetal gonads must be taken into account. In the first weeks, these glands are undifferentiated between male and female fetuses and complete the cellular differentiation process around the 44th post-conceptional day (Baker and Scrimgeour, 1980). The production of testosterone in male fetuses begins during the 2nd month of pregnancy and reaches the maximum in the 2nd trimester (Chung and Auger, 2013), just before when vagal tone starts to rise (Mulkey and du Plessis, 2019).

The different patterns of ANS development could also be due both to a different sex-specific susceptibility to environmental stimuli and a different neurodevelopmental trajectory of male and female fetuses, which could lead to sex-specific developmental intervals of maximum susceptibility to environmental exposures (Buss et al., 2009; Bale, 2016).

Moreover, considering the role of the vagus nerve in the CAP, which involves immune cells such as macrophages (Borovikova et al., 2000; Tracey, 2007; Fairchild et al., 2010; Frasch et al., 2016), and the sex-related neurodevelopment impact of microglial functions (Hanamsagar and Bilbo, 2016), the different characteristics of the developmental pattern of vagus nerve between male and female fetuses could predispose for sex-specific susceptibility and outcomes in many neurodevelopmental disorders (Berntson and Cacioppo, 2004).

### 3.1.2. Maternal factors

#### 3.1.2.1. Maternal distress

The definition of stress has evolved through time, but it is now accepted that a stressor is an uncontrollable environmental demand that exceeds the physiological imbalance of the organism (Selye, 1950; Koolhaas et al., 2011). Several uncontrollable situations such as natural disasters, relational or financial problems, bereavement and/or stressful daily hassles can threaten a woman's life during her pregnancy. These situations increase the risk of impairment of fetal brain development resulting in emotional, behavioral and/or cognitive problems in later life (Glover, 2015; Antonelli et al., 2021).

Historically, research findings oriented to understand the mediators that underlie the basis of fetal programming pointed out the HPA axis disregarding the role of ANS. However, more recently, several papers brought to light the importance of the ANS highlighting the fact that the early developmental disruption of the connections between the ANS and the limbic system might restrict the capacity of the ANS to respond to the environment and can be later implicated in neuropsychiatric disorders of the infant (Montagna and Nosarti, 2016; Frasch et al., 2018a, 2020b). Since, according to the Polyvagal theory (Beauchaine et al., 2007), the social/emotional development is related to the vagal system, any insult such as maternal distress suffered during the autonomic trajectory development will interfere with the maturation of the cerebral cortices structures and of the ANS disrupting the development of the Social Engagement System of the child (Porges and Furman, 2011).

Moreover, an impairment in vagal balance, particularly a decreased tone, might be implicated in depression, anxiety, post-traumatic stress disorder, and schizophrenia (Thayer and Brosschot, 2005). We have recently demonstrated that maternal stress affects the coupling of maternal heart rate to fHR showing that the fetuses of stressed mothers show significant decreases in fHR, probably representing a fetal stress memory that may serve as a novel non-invasive biomarker (Lobmaier et al., 2020).

### 3.1.2.2. Maternal pain

Maternal pain during pregnancy is a symptom that should be seriously taken into account by clinicians, since it could underlie severe pathological conditions, which, in turn, can lead to negative outcomes, both for the mother and the fetus (Brown and Johnston, 2013). Many pathologies, even pre-existing before conception, may induce an acute exacerbation of the chronic pain during the pregnancy, including rheumatoid arthritis, sickle cell disease, Ehlers-Danlos syndrome, and vulvodynia, tuberous sclerosis, and irritable bowel syndrome (Ray-Griffith et al., 2018).

The most frequently observed symptoms are: low back pain, headache, pelvic girdle pain (PGP), leg cramps, breast tenderness, abdominal pain, and ligament pain (Davis, 1996; Jarrell, 2017; Lutterodt et al., 2019). Some symptoms, such as PGP and low back pain, could be physiological consequences of the musculoskeletal adaptations of the maternal body, due to an increasing uterine volume (Davis, 1996; Vermani et al., 2010; Casagrande et al., 2015; Sehmbi et al., 2017; Lutterodt et al., 2019). This could be a potential causing factor of the time-dependent onset of pain symptoms during the pregnancy.

Continued pain and discomfort throughout the pregnancy can induce both a sensitization of the maternal nervous system, with a lower pain threshold (Gintzler, 1980; Cogan and Spinnato, 1986; Eid et al., 2019), and a dysregulation in the inflammatory response, mediated by the CAP via the vagus nerve (Garzoni et al., 2013).

The patterns of cytokine response and the neuroimmune physiology of the fetus and the placenta mirror the maternal ones throughout pregnancy (Sherer et al., 2017; Prins et al., 2018; Sherer et al., 2018; Peterson et al., 2020a). These patterns are altered by pain and could hence not only be pre-emptive for negative outcomes for labor and birth (Goldenberg et al., 2008), but also suggest their influence on the neurodevelopment of the fetal vagal system itself (Borovikova et al., 2000; Tracey, 2009; Frasch et al., 2016).

The maternal pain perception, moreover, can undermine her emotional state, contributing to increased stress and worries (Persson et al., 2013; Clarkson and Adams, 2018; Mackenzie et al., 2018; Lutterodt et al., 2019), and this could additionally interfere with the vagus nerve development of the fetus.

Despite the evidence of positive effects of non-surgical and non-pharmacological treatments, such as physiotherapeutic interventions, acupuncture, or osteopathic manipulations (Gutke et al., 2015; Gallo-Padilla et al., 2016; Ray-Griffith et al., 2018; Smith et al., 2018; Cerritelli et al., 2020), the use of drugs for pain relief (e.g., low-dose aspirin, acetaminophen, non-steroidal anti-inflammatory drugs, and opioids) during pregnancy is still very common (Ray-Griffith et al., 2018). These therapies can affect fetal neurodevelopment: the increase prescriptions for opioids during pregnancy, for example, correlates to an increased incidence of neonatal abstinence syndrome (NAS), resulting in sleeping and feeding difficulties, as well as a dysregulation in newborns' CNS (Pryor et al., 2017; Ray-Griffith et al., 2018). Hence, further research is needed about these drugs' impact on fetal vagus nerve development.

### 3.1.2.3. Maternal nutrition, microbiota, and metabolic pathologies

Maternal nutrition is a substantial part of the "developmental or metabolic programming" of newborns (Koletzko et al., 2018, 2019), with short- and long-term consequences on health and a relevant impact on fetoplacental growth patterns (Morrison and Regnault, 2016).

The inadequate intake or absorption of nutrients, such as polyunsaturated fatty acids, folic acid, and vitamin B12, may result in altered fetal neurodevelopment (Starling et al., 2015; Wang et al., 2015), and lead to important consequences on neurobehaviour during infancy (Gernand et al., 2016; Bordeleau et al., 2021). Maternal malnutrition has an impact also on the endocrine system development: the fetus is overexposed to maternal cortisol, due to the lack of placental control systems (Allen, 2001; Seckl and Meaney, 2004), and the fetal HPA axis can be impaired, possibly leading to a dysregulation of sympathovagal balance and inflammatory set point (Christian and Stewart, 2010).

Even if further research is needed, maternal vitamin D intake deficit correlates with altered metabolic and contractile development of the heart and the control mechanisms of blood pressure in fetuses (Morris et al., 1995; Gale et al., 2008). Similarly, calcium, zinc and iron assumption during pregnancy seems to affect fetal cardiovascular development and autonomic regulatory function (Christian and Stewart, 2010). Therefore, it may be assumed that vagus nerve development itself could be affected by maternal malnutrition.

Inadequate maternal nutrition could also lead to excessive weight gain and gestational diabetes, with negative outcomes for the fetus (AlSaif et al., 2015; Edlow, 2017; Peterson et al., 2020b; Tong and Kalish, 2020). Maternal obesity, in particular, both before conception and during pregnancy, is associated with impaired fHRV and poorer ANS development (Christifano et al., 2020).

Maternal metabolic dysregulation seems to be related to gut dysbiosis (Hasain et al., 2020; Tong and Kalish, 2020) and, thanks to the direct link due to the feto-placenta-maternal unit (Nuriel-Ohayon et al., 2016; Laursen et al., 2017; Bordeleau et al., 2021), it could negatively influence the development of fetal microbiota (Zhou and Xiao, 2018; Tong and Kalish, 2020). Maternal metabolic dysfunctions and dysbiosis can affect the systemic inflammatory response of the mother and the fetus (Hasain et al., 2020; Peterson et al., 2020b). Moreover, considering the correlation between the gut-brain-microbiome axis and immune system via neuroendocrine signalling (Aidy et al., 2015; Vitetta et al., 2018; Fuhler, 2020), these maternal affections can affect fetal neurodevelopment and potentially lead to adverse health outcomes later in life (Borre et al., 2014; Abdel-Haq et al., 2018; Tong and Kalish, 2020; Bordeleau et al., 2021).

Lastly, this feto-maternal pro-inflammatory state could interfere with the set point of the fetal vagal regulatory system through CAP dysregulation, thus impairing fetal intestinal permeability and integrity (Garzoni et al., 2013).

### 3.1.2.4. Circadian clock and vagal activity

During the 3rd trimester of pregnancy, emerging fetal sleep states are detectable, with the onset of the functional organization of the sleep cycle around 28-30 WGA, and these patterns gradually consolidate around 36 WGA (Van den Bergh and Mulder, 2012). At this developmental stage, the fetus shows alternating behavioural states, passing through non-REM sleep and REM sleep, which lasts about 70–90 minutes (Visser et al., 1992), with usually a wake state duration under 10%.

Sleep cycles are essential for fetal neurodevelopment (Graven and Browne, 2008). The deprivation of the REM sleep phase during uterine life can lead to altered cortical and brainstem development and to the supersensitivity to the noradrenaline of the pyramidal cells in the hippocampus, impairing the receptors' sensitivity (Mirmiran et al., 2003). Consequently, the fetal and infant capacity to homeostatically adapt to the environment could be impaired (Van den Bergh and Mulder, 2012).

fHR patterns, together with GMs and REM recording, are the main factors to define the passage through fetal sleep states (Van den Bergh and Mulder, 2012), pointing out the direct relationship between vagal activity and circadian rhythm.

Although the ANS and vagus nerve activity appear to have an endogenous circadian rhythm (Hilton et al., 2000; Kentish et al., 2013), with a pacemaker system that regulates its behaviour regardless of sleep-awake states (Malpas and Purdie, 1990; Hilton et al., 2000), a circadian variation pattern of vagal activity can be recorded via HR variations.

The circadian dynamics of fHR is synchronized with the maternal rhythm of the activity at rest, heart rate, cortisol, melatonin and body temperature. It is possible, therefore, to assume that the mother, considering the limited light-dark variation during intrauterine life, entrains the developing circadian rhythm of the fetus to this cycle (Mirmiran et al., 2003). In this sense, dysregulations in mother circadian rhythm, due to disturbed sleep-awake patterns, could interfere with the development of circadian control systems of the fetus and, consequently, with the circadian profile of the vagus nerve, impacting the physiologic vagal neurodevelopment itself.

Several factors can normally affect the sleep states of the mother during pregnancy: the hormonal, biomechanical, and behavioural changes that occur during the three trimesters, often lead to increased sleep disturbances, particularly in the last trimester (Sahota et al., 2003; Pien and Schwab, 2004; Facco et al., 2010; Mindell et al., 2015).

It has been reported in these cases a correlation with deficit in fetal growth, adverse outcomes for labour and birth, and even preterm birth (Lee and Gay, 2004; Micheli et al., 2011; Zafarghandi et al., 2012; Warland et al., 2018). Especially in the 3rd trimester, it is frequently reported the onset of sleep-disordered breathing (SDB) (Santiago et al., 2001; Edwards et al., 2002; Pien and Schwab, 2004; Venkata and Venkateshiah, 2009). SDB can disturb not only the physiological patterns of maternal sleep, but it could also lead to nocturnal hypoxia, which produces an enhanced activation of the maternal ANS (Pien and Schwab, 2004; Sahin et al., 2008). There are also repercussions on the fetus, in response to these apneic events: fHR decelerations and decreased breathing movements, which could affect neonates' neurobehavior, even if currently there are not enough data to support definitive conclusions about long-term outcomes (Pien and Schwab, 2004).

Lastly, maternal sleep deprivation seems to be related to the depressive state (Jarczok et al., 2018) and leads to increased systemic inflammation, with the increment in the production of pro-inflammatory serum cytokines (Chang et al., 2010). In this context, further research is needed to better clarify the relationship between maternal neuroendocrine dysregulation, due to chronobiological impairments, and the impact on fetal neurodevelopment, particularly on the CAP.

### 3.1.2.5. Maternal diseases

Maternal pathologies influence the onset of altered neurodevelopment patterns during fetal life, which could lead to actual neurologic disorders during childhood (Faa et al., 2016; Vohr et al., 2017). In particular, metabolic and cardiovascular pathologies, like obesity, diabetes, hypertension, and preeclampsia, can disrupt vagus nerve development (Brown et al., 2008; Russell et al., 2016; Moors et al., 2020).

The pathologies that induce chronic hypoxemia in the mother, including SBD (Pien and Schwab, 2004) and anemia, may induce dysregulations in the vagal control systems. Such hypoxic states can also be secondary to infections, including chorioamnionitis, leading to several hemodynamic abnormalities and cardiovascular compromise (Frasch, 2018). The endotoxemia, through lipopolysaccharide molecules, seems to interfere with the cardiac pacemaker synchronization system, altering the fHRV pattern (Frasch and Giussani, 2020).

Many viral infections, such as cytomegalovirus and other TORCH organisms or Zika virus (ZIKV) infection, can lead to preterm birth and induce severe neurodevelopmental alterations in the fetus (Silasi et al., 2015; Vohr et al., 2017). These pathologies are associated with fetal inflammatory response syndrome, which manifests as increased pro-inflammatory markers, including IL-6, IL-1, and TNF-α (Gotsch et al., 2007). As a result, oligodendrocytes and their progenitors are damaged during a critical period of brain development, thus altering brain myelination and leading even to periventricular leukomalacia (Kadhim et al., 2001). The increase in cytokine production consequent to the inflammatory response can enhance the risk for cardiovascular disease, impairing CAN, autonomic reactivity (Harrison et al., 2013), and, therefore, affecting the vagal development. Moreover a recent pilot study described that in utero ZIKV-affected babies display HRV metrics abnormalities. Even early prenatal exposure to the virus, indeed, seems to create a chronic hypoxic environment to the fetus, affecting vagus nerve developmental patterns and imprinting postnatally the ANS cardiac pacemaker activity (Herry et al., 2021).

The implications of these infection-associated immunological events on fetal neurodevelopment seem to be time-dependent, thus supporting the clinical relevance of carefully evaluating the interfering factors throughout the entire pregnancy (Meyer et al., 2007).

Similarly, even autoimmune diseases can involve a dysregulation in maternal inflammatory response (Gordon, 2004; Lu-Culligan and Iwasaki, 2020; Han et al., 2021). Systemic lupus erythematosus and antiphospholipid antibody syndrome, for instance, entail the presence of IgG antiphospholipid antibodies, which can cross the placenta, interfering with the brain development and impacting the cardiovascular system of the fetus (Buyon, 1998; Nalli et al., 2017). These pathologies can dysregulate the neuroendocrine axis, altering, in particular, the cortisol production (Wilder, 1998; Sheng et al., 2020) and, thus, affecting sympathovagal balance development.

Even conditions that induce secondary hypertension or HPA dysregulation, such as pheochromocytoma, hyperaldosteronism, or Cushing syndrome, can affect fetal neurodevelopment (Morsi et al., 2018; Corsello and Paragliola, 2019).

Lastly, maternal mood disorders and depression seem to strongly impact the fetal neuroendocrine systems, in particular the amygdala (McEwen et al., 2016) and the hippocampus (Nemoda et al., 2015), strongly involved in the CAN (Harrison et al., 2013; Thome et al., 2017). Moreover, maternal psychopathology influences HRV tracking also after birth, showing a higher mean HR and lower vagal modulation and confirming the impact of this prenatal disorder on vagal neurodevelopment (Dierckx et al., 2009).

### 3.1.2.6. Nicotine/alcohol exposure

Smoke and alcohol exposure can induce in premature newborns higher sympathetic function, lower parasympathetic function, and had less cardiac autonomic adaptability (Mulkey and du Plessis, 2019) During the 1st trimester, there are significant negative correlations between the amount of maternal cigarette smoking and fHRV metrics, highlighting the disrupted temporal organization of autonomic regulation before birth (Zeskind and Gingras, 2006; Kapaya et al., 2015).

Prenatal nicotine exposure (PNE) is associated with higher systolic blood pressure and altered HRV after birth (Nordenstam et al., 2019). Furthermore, the diminished vagal cardiovascular capacity to respond to hypoxic and hypercapnic states in the fetus, due to the abolishment of the serotonergic neurotransmission to cardiac vagal neurons caused by PNE, has been hypothesized as a likely link to SIDS (Kamendi et al., 2006, 2009).

The effects of prenatal alcohol exposure (PAE) on fetuses are collectively known as fetal alcohol spectrum disorders (FASD). Of the various organ systems affected by FASD, the brain is the most severely impacted: even if the hippocampus seems to be spared, other CAN structures show an impaired development, including the thalamus, hypothalamus and prefrontal cortex, and also the spinal cord (Caputo et al., 2016).

Alcohol easily crosses the placenta and can disrupt maternal-fetal hormonal interactions, leading to long-term impairments on neuroendocrine and immune competence (Zhang et al., 2005). PAE also impacts the limbic-HPA axis development and functioning, leading to disrupted cortisol response (Ouellet-Morin et al., 2011). PAE, therefore, induces a dysregulated fetal immune response, increasing the risk for infections, chorioamnionitis, and/or placental abruption, which can favor premature delivery (Reid et al., 2019). The consequences on the long-term capacity of CAP inflammatory response are worthy of further investigations.

Lastly, early pain reactivity measured through HRV in alcohol-exposed newborns seems to be blunted, suggesting different responsiveness to environmental stimuli even after birth (Oberlander et al., 2010). This seems to confirm the PAE effects on HPA axis stress reactivity, possibly via changes in central serotonin levels and/or HPA axis functions, and on the vagus nerve development itself (Haley et al., 2006).

### 3.1.3. Placenta

#### 3.1.3.1. Placental dysfunctions

A growing body of evidence suggests the placenta's role in modulating fetal development. The placenta can be defined as a neuro-immune-endocrine secretive organ, with the essential role of conveying nutrients and developing modulatory signals to fetuses, limiting their exposure to any factors that could alter their physiologic developmental patterns.

Many factors can affect placental physiology, such as maternal nutrition, nicotine or alcohol exposure, drugs, pathologies, and stress (Nugent and Bale, 2015). This results in an overexposure of the fetus to maternal cortisol, that in turn may produce an overactive fetal cortisol response and may be a predisposing factor to develop disorders related to elevated HPA axis activity, which could lead to impairments in environmental stimuli adaptability by ANS in newborns (Nugent and Bale, 2015).

The dysregulation of placental functions can induce an insufficient intake of nutrients for the fetus, thus, especially during mid and late gestation, impairing the development of various brain areas, including the hypothalamus that seems to be particularly sensitive to the nutrient deficiency (Nugent and Bale, 2015).

Moreover, through the production of corticotropin-releasing hormone (CRH), the placenta stimulates adrenocorticotropic hormone release in the anterior pituitary of the fetus, affecting both fetal HPA axis and amygdala development (Seckl and Meaney, 2004). A dysregulation in CRH production, secondary to various maternal conditions, such as hypoxia, increased inflammatory cytokines and glucocorticoids, stress, preeclampsia, and eclampsia, may lead to altered neurodevelopmental patterns of these fetal structures (Allen, 2001; Bale, 2016; Sheng et al., 2020). It could also affect the role of placenta as the immune pacemaker of pregnancy, increasing the risk of preterm birth (Allen, 2001; Peterson et al., 2020a) and thus impairing vagal development during the last weeks of gestation.

Recently, placenta calcifications samples collected at delivery have been speculated to act as a memory of prenatal maternal stress and diseases exposure, hence highlighting the potential role of placenta in regulating the offspring's future cardiovascular and metabolic health, even through the in utero modulation of the developing ANS (Wallingford et al., 2018).

In addition, even mild inflammatory states in the mother might affect the placental conversion of maternal tryptophan to serotonin upstream of the fetal brain (Goeden et al., 2016). This affection ultimately interferes with cerebral fetal neurodevelopment, including the nucleus accumbens and hippocampus, that are involved in CAN (Seckl and Meaney, 2004).

In conclusion, many diseases can impact the placental function, which in turn seems to have a role in regulating the neurodevelopment of the main actors that interact with the fetal vagus nerve, and the monitoring of placenta secretome could be a useful biomarker of diseases and potential developmental disruptions of the ANS in the fetus (Aplin et al., 2020).

### 3.1.3.2. Placental microbiome

Although the womb was usually considered sterile, recently it has been suggested that the placenta harbors a unique microbiome, composed of nonpathogenic commensal microbes (Aagaard et al., 2014). The placental microbiota seems to display an evolutionary pattern during pregnancy and appears to be diversified in the various areas of the placenta (Aagaard et al., 2014; Cao et al., 2014; Pelzer et al., 2017).

Many authors substantiated the materno-fetal transmission of the microbiome through the placenta during prenatal life, identifying the presence of the microbiota both in the peripartum placenta (Zheng et al., 2015; Pelzer et al., 2017) and in the meconium (Walker et al., 2017), partially regardless of the delivery mode.

Placenta dysbiosis can be related to antepartum infections (Aagaard et al., 2014; Doyle et al., 2017; Pelzer et al., 2017), maternal nutrition (Aagaard et al., 2014), maternal weight gain (Antony et al., 2015; Zheng et al., 2015; Benny et al., 2019), gestational diabetes mellitus (Bassols et al., 2016; Pelzer et al., 2017), use of probiotics/antibiotics (Pelzer et al., 2017) and periodontal pathogens (Fischer et al., 2019), and can lead to several adverse outcomes, such as preeclampsia and chorioamnionitis (Amarasekara et al., 2015; Doyle et al., 2017; Fischer et al., 2019; Olaniyi et al., 2020).

Moreover, it has been observed a correlation between placenta dysbiosis and preterm birth (Aagaard et al., 2014; Antony et al., 2015; Bassols et al., 2016; Prince et al., 2016; Pelzer et al., 2017; Fischer et al., 2019). More than the infections themselves, it would seem it is the dysregulation of the placental inflammatory response that could affect the production of placental CRH and stimulate the fetal HPA axis, leading to preterm labour and birth (Parris et al., 2021). The high immunoreactivity of the premature infants' gut could also lead to neurodevelopmental disorders (Cao et al., 2014).

The fetoplacental transmission of the placental microbiome may occur through the ingestion of the amniotic fluid and via umbilical cord blood supply (Cao et al., 2014; Walker et al., 2017). The matches collected between the placental and fetal gut and lung microbiomes seem to be associated with neonatal airway complications and enterocolitis (Al Alam et al., 2020; Parris et al., 2021), possibly affecting the ANS-mediated control mechanisms of these systems. Since the placental microbiome could also affect metabolic pathways, thus leading to a pro-inflammatory environment (Gomez-Arango et al., 2017; Fischer et al., 2019), it could also affect the ANS and vagus development of the fetus.

Despite this evidence, other authors claim that placental microbiome does not exist – the aforementioned results would be just due to methodological errors – or that there is only extremely low biomass, probably leading to minor clinical effects (Lauder et al., 2016; Leiby et al., 2018; Kuperman et al., 2020; Parris et al., 2021).

Nevertheless, the placental microbiome could be an interesting factor to be taken into account by the clinicians as a potential regulator of vagus nerve development.

### 3.1.4. Exogenous (non-maternal) factors

#### 3.1.4.1. Environmental stimuli

The variations in ambient and core temperature appear to impact newborns HRV (Massaro et al., 2017): in particular, extreme variations seem to affect the autoregulatory capacities of the cardiac system and, consequently, HRV, possibly due to a dysregulated release of excitatory neurotransmitters (Tsuda et al., 2018). Particularly warmer- or colder-than-average temperatures seem to be associated with increased systemic inflammation and alteration in placental blood flow (Martens et al., 2019; Sun et al., 2019), leading to dysregulation of metabolic and immune substrates of the fetus, and possibly influencing the regulatory functions of ANS.

Exposure to acoustic stimuli, through vibroacoustic stimulation (VAS), is one of the most used benchmarks for the assessment of fetal distress, since VAS induces a somato-cardiac effect mediated by the fetal behavioural states (Busnel et al., 1992). This cardiac-orienting response does not seem to be detectable before the functional maturity of the fetal autonomic system, which occurs around 32 WGA (Krueger and Garvan, 2019). The fetal responsiveness recorded via fHRV patterns after VAS stimulation (Abrams and Gerhardt, 2000) is refined as the pregnancy proceeds, being a useful parameter to evaluate the maturation of ANS and vagus nerve (Buss et al., 2009). As the nervous system of the fetus develops, around 36-39 WGA a different fHRV pattern is detectable according to different acoustic stimuli, as for example human speech (Lecanuet et al., 1992; Krueger and Garvan, 2019).

Even if the light-dark cycle does not seem to significantly affect the fetal circadian rhythm, there are age-dependent fHRV patterns after transabdominal fetal stimulation with halogen light, similarly to the VAS (Peleg and Goldman, 1980), which seem increasingly definite in the later stages of pregnancy, when the fetal optic pathways and ANS control on the cardiac activity reach maturity (Thanaboonyawat et al., 2006).

The persistent exposure to photic stimuli or high noise during pregnancy, for instance, due to the mother's occupation, seems to correlate with reduced fetal growth (Selander et al., 2019), increased incidence of congenital heart diseases (Gong et al., 2017), impaired development of cortical areas (Zhang et al., 1992; Wilson et al., 2008), and potential disruption of the vagus nerve maturation.

The occupational and environmental exposure to pollution may be another important disrupting factor for fetal ANS development. A growing body of evidence suggests that chemicals exposure can impact fetal growth (Dejmek et al., 1999; Snijder et al., 2012; Desrosiers et al., 2015) and is a risk factor for congenital heart diseases (Gong et al., 2017), preterm birth and severe neurodevelopmental abnormalities (Zhang et al., 1992; Lacasaña et al., 2006; Langlois et al., 2012; Yurdakök, 2012). Even if the effects on the vagus nerve are still to be defined, the hazards exposure may affect the fetal motor activity (DiPietro et al., 2014) and be related to delayed neurobehavioural development in neonates (Handal et al., 2008). It is instead well known that methylmercury from seafood impairs the higher centres of cardiac autonomic function (Karita et al., 2018), leading to severe ANS abnormalities (Sørensen et al., 1999; Grandjean et al., 2004; Murata et al., 2006; Gribble et al., 2015).

Lastly, even electromagnetic field (EMF) radiations during pregnancy, due to overexposure to mobile phones, diagnostic instruments, or occupation, seem to be a risk factor for preterm delivery (Roşu et al., 2016) and to have effects on ANS development (Rezk et al., 2008).

### 3.1.4.2. Drugs during pregnancy and labor

In the course of pregnancy, mother and fetus could undergo a variety of drug administrations, many of which can cross the placenta affecting fetal development in several ways (Miljković et al., 2001).

*Drugs to support fetal growth*

Synthetic glucocorticoids (sGCs) are the therapy of choice to support fetal maturation for the risk of preterm delivery. Usually, it is administered between 24 and 34 WGA, mainly to ensure optimal fetal pulmonary development (Mulder et al., 2004; Committee on fetus and newborn and section on anesthesiology and pain medicine, 2016).

As excessive maternal cortisol, sGCs override the placenta mediator system (Sheng et al., 2020) and induce transient HRV modification toward sympathetic suppression (Multon et al., 1997; Senat et al., 1998; Subtil et al., 2003; Mulder et al., 2004; Schneider et al., 2010). These modifications seem to last at worst 4 days after the first dose (Mulder et al., 2004; Verdurmen et al., 2013) and the influence on ANS modulation seems minor (Verdurmen et al., 2018; Noben et al., 2019). However, the long-term impact is currently under investigation. There appears to be an association between multiple courses of sGCs prenatal administrations and infant neurodevelopmental abnormalities (Spinillo et al., 2004). Nevertheless, the presence of many confounding factors and the small sample sizes make unclear the mechanism of action of glucocorticoids on fHRV and consequently the impact on ANS development (Verdurmen et al., 2013).

Prenatal exposure to sGC may interfere both with the fetal endogenous cortisol sensitivity, which involves the hippocampus, amygdala, and HPA axis and with the fetal cardiovascular and metabolic regulatory systems (Seckl and Meaney, 2004). This could lead to profound epigenetic changes in the fetal nervous system development and specifically in the vagal functionality, with possible long-term consequences (Crudo et al., 2013; Chang, 2014). This in utero programming effect on the fetal HPA axis could last until adulthood: there are recent observations of altered R-R interval variability in adult offspring exposed to elevated fetal sGCs (Sheng et al., 2020), a result that deserves further investigations to better understand the sGCs potential role in neurodevelopmental altered pathways.

*Drugs for maternal pathologies and drug abuse*

Most of the medication administered to the mother during pregnancy can cross the placental barrier and reach the fetus.

Selective serotonin reuptake inhibitors (SSRIs) seem to induce alterations in the fetal limbic system (Lattimore et al., 2005), disrupting the central serotonin signaling (Oberlander et al., 2009). Exposure to SSRIs during the 3rd trimester is associated with blunted FMs and reactivity to VAS, as well as with reduced fHRV at 36 WGA (Oberlander et al., 2009; Nguyen et al., 2019). The fetuses, moreover, appear to have an increased risk of preterm birth (Morrison et al., 2005; Suri et al., 2007; Oyebode et al., 2012) and poor neonatal adaptation syndrome (Lattimore et al., 2005; Sivojelezova et al., 2005).

Neonates affected by in utero exposure to SSRIs display fewer HRV rhythms and changes in behavioral states, based on the time drugs were administered (Zeskind and Stephens, 2004). They may show altered pain reactivity and parasympathetic cardiac modulation during recovery after an acute noxious event (De las Cuevas and Sanz, 2006), which lasts up to two months, and also altered HPA stress response patterns and hippocampal plasticity (Morrison et al., 2005; Oberlander et al., 2009; Gemmel et al., 2019).

Among antihypertensive drugs, methyldopa does not seem to particularly affect fetal wellbeing. In contrast, the first-trimester exposure to angiotensin-converting enzyme inhibitors is associated with the greater incidence of malformations of the cardiovascular and central nervous system (Podymow and August, 2008, 2011), and the timing of exposure seems to be particularly relevant (Caton Alissa R. et al., 2009). The use of nifedipine concurrently with magnesium sulfate ($MgSO_4$) seems associated with severe hypotension, neuromuscular blockade, and cardiac depression in the fetus (Khedun et al., 2000).

The effects of prenatal opioids exposure on the fetus are widely described. Although there are not evident depressive effects on non-stress test reactivity or variability in the case of daily opioid use for chronic pain (Brar et al., 2020), the current guidelines recommend a prescription at the lowest effective dose (Ray-Griffith et al., 2018).

The opioid prenatal overexposure has instead neurotoxic consequences on several fetal systems: cardiovascular, respiratory, neurobehavioral, metabolic, and neuroendocrine (Conradt et al., 2018). Early exposure of buprenorphine induces higher levels of fHRV, whereas at 32-36 GWA fetuses display less suppression of motor activity (Jansson et al., 2011; Conradt et al., 2018), indicating the involvement of the vagus nerve system. Opioid dependence, moreover, is associated with the increased risk for neonatal opioid withdrawal syndrome (NOWS) (McCarthy et al., 2017; Conradt et al., 2018, 2019).

After birth neonates with NOWS show increased levels of norepinephrine, corticotropin, noradrenaline, and acetylcholine, and lower dopamine and serotonin levels (Conradt et al., 2018), thus suggesting a possible disruption in vagal and CAP activity. Exposure to opioids prenatally could result in programming effects on stress response systems, leading also to long-term consequences (Conradt et al., 2018, 2019).

Opioids are often used as analgesic therapy during labor. There are many observed effects of this medication on fHRV and fetal breathing patterns, mostly transient and dose-/time-dependent (van Geijn et al., 1980; Kariniemi and ÄMmälä, 1981; Low et al., 1981; Viscomi et al., 1990; Capogna, 2001; Mattingly et al., 2003; Shekhawat et al., 2020). To better understand the real implications of intrapartum opioids administration on fetal vagus nerve development, however, it is essential that future research distinguishes its effects from that produced by other substances and associated environmental stressors, illustrating the role of the timing of exposure in specific windows of fetal neurodevelopment, as well as the potential long-term outcomes (Conradt et al., 2018, 2019).

It is also in common use during the peripartum period to administer the mother antibiotic prophylaxis. This seems to be associated with an increased risk for antibiotic-resistant neonatal sepsis (Mercer et al., 1999; Hantoushzadeh et al., 2020). Moreover, prophylactic antibiotic use during vaginal or cesarean delivery may indirectly lead to epigenetic changes in the fetus, with possible implications in the immune system and stress response (Tribe et al., 2018), consequently involving vagus nerve development and suggesting, therefore, the utility to consider potential alternative approaches (Ledger and Blaser, 2013).

*Drugs used to manage the timing of delivery*

MgSO$_4$, which is often administered between 23 and 27 WGA or intrapartum for its tocolytic effect to delay labor onset in pregnancy at risk for preterm birth (Ramsey and Rouse, 2001; Ingemarsson and Lamont, 2003), does not seem to show a significant change in fHRV (Stallworth et al., 1981; Nensi et al., 2014). These results do not vary in co-administration with opioids (Cañez et al., 1987), whereas the association with sGC should deserve greater attention by the clinicians (Verdurmen et al., 2017). Moreover, the prolonged administration of MgSO$_4$ seems to be associated with a change in fetal breathing movements (Peaceman et al., 1989) and to affect fHRV dose-dependently (Cardosi and Chez, 1998; Kamitomo et al., 2000).

## 3.2. The labor/delivery influence on the ANS

Labor is divided into three stages. The 1st stage begins when significant and regular contractions start and ends when there is a full cervical dilation of 10 centimeters. The 2nd stage of labor begins with complete cervical dilation and ends with the delivery of the fetus. The 3rd stage commences when the fetus is delivered and ends with the delivery of the placenta (afterbirth) (Hutchison et al., 2021).

In this section, we will review the influences that the events that can occur in the first two stages of labor can have on the development of the vagus nerve and ANS. In particular, we will study what events can interfere with the correct development of these two systems by analyzing the evidence from preclinical and clinical studies in this regard, with the aim of preventing potential complications with long-term consequences.

Changes in fHR and fHRV during labor can reveal alterations in fetal reserve to survive the trial of labor and in ANS development, specifically, also in vagus nerve modulation (Chiera et al., 2020). However, it is important to highlight that so far fHR monitoring has not been enough to prevent fetal injury during labor and, consequently, long-term complications. In fact, the role of fHR monitoring (electronic fetal monitoring, EFM, to be precise), despite being used in over 90% of hospitals during delivery, remains controversial (Schifrin, 2020). There is no clarity on its usefulness in decreasing perinatal mortality or cerebral palsy when measured by a CTG which has been explained by the incorrect focus of EFM on prediction of acidemia, a poor correlate to fetal injury, instead of predicting fetal cardiovascular decompensation per se as well as the limitations of CTG technologies in capturing the short-term time scale fluctuations of HRV reflecting vagal modulations (Alfirevic et al., 2017; Frasch et al., 2017; Frasch, 2018; Gold and Frasch, 2021). Consequently, EFM fails to identify fetuses at risk of brain injury; this is because there is no clear correlation between fetal brain injury and acidemia (Westerhuis et al., 2007; Cahill et al., 2017). Additional modalities of EFM have been proposed such as intrapartum maternal transabdominal electrocardiography (ECG) and electroencephalography (EEG) which will increase the ability of EFM to predict fetal brain injury. Fetal EEG, while feasible, has not yet made its way into clinical practice. Fetal scalp electrodes can also be used to directly measure pH during childbirth, but this technique, being even more invasive, is not yet fully accepted and also has some complications (Sabir et al., 2010).

### 3.2.1. Preclinical studies

Through preclinical studies on animals and using fHRV, which remains the best method to monitor the fetus, researchers are trying to find solutions to improve and increase the prevention of specific complications. The preclinical studies analyzed in this review were selected from among those carried out on the ovine species. Sheep are in fact an important model for simulating human pregnancy. Research on the ovine species, reproducing the conditions of human labor, is indeed leading to the discovery and evidence of different algorithms to prevent pathological complications (Frasch et al., 2011; Wang et al., 2014).

In this paragraph, we will consider the studies that simulate the 1st stage of labor.

During this stage, uterine contractions expose the fetus to continuous and significant hypoxic stress. Perinatal hypoxia contributes significantly to the possibility of perinatal brain injury (Gotsch et al., 2007). Statistically, severe fetal acidemia (pH <7.00) has been shown to occur in between 0.5% and 10% of human births. Of the neonates in this series, 20% will have neurological consequences including cerebral palsy and HIE (Goldaber et al., 1991; DuPont et al., 2013).

Among the possible fHRV metrics, RMSSD is a metric that allows researchers to assess what influence labor may have on the vagus nerve. Indeed, it is known to increase with worsening acidemia (Task Force of the European Society of Cardiology and the North American Society of Pacing and Electrophysiology, 1996; Frasch et al., 2007, 2009a, 2009b; Durosier et al., 2014).

Gold et al. hypothesized, through the use of an algorithm based on RMSSD, the possibility of identifying the precise moment, during labor, when fetuses begin to lose maintenance of cardiac output, that is, when they show a hypotensive pattern of pathological arterial blood pressure (ABP). This hypothesis is highly clinically relevant because, in the case of repetitive umbilical cord occlusions with worsening acidemia, fetuses may easily develop cardiovascular decompensation, a phenomenon that occurs through ABP responses to HR decelerations. Such a failure in maintaining cardiac output during labor could facilitate brain injury in newborns and can be detected on the individual basis using fHRV analysis from EFM (Gold et al., 2021; Roux et al., 2021).

The above-mentioned studies show us what impact labor can have on the ANS and the vagus nerve. In fact, the brain, through the CAN, controls the sympathetic and parasympathetic preganglionic motoneurons and regulates the activity of the vagus nerve (Benarroch, 1993). Consequently, a cerebral alteration, caused by a brain injury, affects the ANS including the vagus. This is the reason why a measure of vagal modulation like RMSSD could help predict a similar event.

Another common complication during labor is given by cardiac repercussions that can affect the fetus. Hypoxia can induce heart failure which is revealed by accelerations and decelerations in fHRV (Rivolta et al., 2014). This event shows how the fetal ANS quickly reacts in the face of a stressor. Further research is required to tie these findings with the recent discovery of the intrinsic fHRV produced by the heart's sino-atrial node (Frasch et al., 2020a). The evidence so far shows that the intrinsic fHRV is influenced by the fetal ANS activity and, conversely, also shapes the ANS activity into the postnatal period, at least in part via the the afferent fibers of the vagus nerve (Herry et al., 2021).

Vagal modulation, as revealed by RMSSD, and global ANS functioning, as revealed by entropic HRV metrics, can also be affected by the neuroinflammation in utero that can occur during labor due to short-term hypoxic acidemia. A key role in the control of inflammation is played by the CAP and a functional neuroimmunological link in the fetus has been shown to improve postnatal health and brain injury (Frasch et al., 2016; Liu et al., 2016). CAP also plays a very important role in preventing and ameliorating NEC cases (Garzoni et al., 2013). During labor, especially in cases of prematurity, there are several conditions that can cause NEC: two of these are certainly hypoxia and acidemia (Sharma and Hudak, 2013). Since the vagus is closely related to the intestinal system, severe intestinal inflammation such as NEC can destroy the cells of the vagal nuclei in the brainstem and inhibit CAP (Fritze et al., 2014).

### 3.2.2. Clinical studies

Labor in humans is certainly associated with physiological changes, most of which are regulated by ANS (Sanghavi and Rutherford, 2014; Soma-Pillay et al., 2016). In this paragraph, we will focus on clinical studies done during both the 1st stage of labor and the 2nd stage of labor.

#### 3.2.2.1. The 1st stage of labor

During this stage, various components come into play that create an alternate activity between SNS and PNS (Norman et al., 2011; Musa et al., 2017). For example, the maternal pain and anxiety due to uterine contractions induce an activation of the sympathetic division (Jones and Greiss, 1982), whereas the release of the hormone oxytocin instead activates the parasympathetic division (Gamer and Büchel, 2012).

A possible critical event during the first stage of labor is certainly the risk of cardiac events. By examining cardiac autonomic modulation (CAM), some authors hypothesized that there might be a predominance of activation of the SNS over the PNS. However, SNS hyperactivity during labor is unlikely to manifest like myocardial infarction, as sympathetic CAM increases simultaneously with HRV metrics related to PNS (Musa et al., 2017).

An important factor that is present in several cases of labor is epidural analgesia (Bautista and George, 2020). Its influence on the development of the ANS and the vagus nerve can be understood through a careful examination of fHRV changes. In fact, epidural analgesia itself does not show precise or significant changes in fHRV (Lavin, 1982), but it can interact with other aspects already existing in the mother. Hypotension and uterine hypertonia, for instance, can create significant imbalances in the fHRV during epidural analgesia in labor (Lavin, 1982). Several studies have also shown a major change in fHRV in women given an analgesic agent such as lidocaine (Boehm et al., 1975; Lavin, 1982).

An aspect that should not be underestimated in the administration of epidural analgesia is the position of the mother. fHRV undergoes fewer changes with the use of the full lateral position, compared with the use of wedged supine position (Preston et al., 1993). The reason behind this evidence could be the greater prevention of aortocaval compression given by the full lateral position (Preston et al., 1993).

The 1st stage of labor does not always occur naturally and without complications: consequently, there may be an artificial intervention that causes the induction of labor to increase uterine contractions. Labor is usually induced through the administration of synthetic oxytocin (sOT). As previously mentioned, oxytocin plays an important role in activating the PNS and it has been shown to dampen SNS (Gamer and Büchel, 2012). However, the induction of oxytocin did not reveal cardiac complications in women with heart disease (Dogra et al., 2019).

#### 3.2.2.2. The 2nd stage of labor

Although the induction of the hormone oxytocin has no cardiac repercussions, it can still influence the development of the vagus nerve and the ANS in the last hours of labor. It has been shown that a pattern of hyperactive uterine contractions (often caused by the administration of sOT), during the last two hours of labor, is strongly associated with acidemia and hypoxia at birth and with fHR and fHRV alterations (Freidman and Sachtleben, 1978;

Woodson et al., 1979; Cibils and Votta, 1993; Ladfors et al., 2002; Jonsson et al., 2008; Verspyck and Sentilhes, 2008; Aye et al., 2014). As previously mentioned, acidemia can induce NEC and consequently inhibit the CAP (Garzoni et al., 2013).

The long-term consequences of developing cardiac control systems and ANS are hence under investigation (Frasch and Giussani, 2020). Moreover, sOT infusion may decrease spontaneous oxytocin release (Jonas et al., 2009), elevating days after birth fetal plasmatic oxytocin concentration, not dampening the HPA axis response and increasing the CAN system control activity (Yee et al., 2016).

Exposure to sOT may therefore lead to epigenetic remodeling in the infant (Tribe et al., 2018; Uvnäs-Moberg et al., 2019), and the documented presence in animal models of specific binding for oxytocin in the vagal dorsal motor nucleus during early embryonic life (Tribollet et al., 1989) suggests a strict link between oxytocin and vagus nerve.

Drugs aside, the most relevant aspect to be evaluated during the 2nd stage of labor is certainly the mode of delivery.

The major critical events that can affect vagus nerve and ANS development occur in premature births and delivery by CS at term (Hillman et al., 2012; Tribe et al., 2018; Mulkey et al., 2019).

The case of premature births is probably the most challenging. Preterm newborns have an immature ANS, which has great repercussions on cardiac and respiratory functions (Mulkey and du Plessis, 2018). In infants born before 36 WGA, there is also a 90% incidence of NEC, as the premature gastrointestinal tract has increased permeability (Garzoni et al., 2013).

Although CS delivery is far faster than a vaginal birth modality, the fetus is still subjected to significant stress that can create changes in the development of some systems, including the ANS (Tribe et al., 2018). An important difference should be moreover emphasized between the CS delivery following a period of labor and CS delivery prior to the onset of labor. Infants born with CS delivery following a period of labor have a higher and more developed ANS tone compared to those born with CS without labor (Mulkey et al., 2019). Furthermore, labor pain before CS helps prevent respiratory disorders and avoid umbilical cord acidemia, thus protecting the developing ANS (Senturk et al., 2015).

Delivery without labor or preterm birth is associated with lower cortisol, angiotensin and catecholamines levels compared to labored or full-term delivery. This situation puts newborns at risk of cardiovascular and metabolic dysfunctions, which can then negatively affect the developing brain by inducing inflammation or glucose deficiency (Hillman et al., 2012; Morton and Brodsky, 2016; Mulkey and du Plessis, 2018).

### 3.3. The incubator influence on the ANS

Once born, newborns have to deal with the extrauterine environment that, especially for preterm babies, is constituted by the incubator. Here, environmental factors such as temperature, light, noise, procedures could induce a high amount of stress, which could become toxic, that is, induce severe alterations in the brain and organic development, with long-lasting complications (De Jonckheere and Storme, 2019; Weber and Harrison, 2019).

#### 3.3.1. Temperature

As thermoregulation is one of the most important ANS functions, the temperature can easily affect ANS development. Indeed, warm incubators tend to increase SNS tone and reduce PNS activity, whereas the opposite is true for incubators 2°C colder than newborns' temperature (Franco et al., 2003; Stéphan-Blanchard et al., 2013). Therefore, temperature that constantly deviates from newborns' one can lead to excessive ANS activation. Indeed, extreme variations of environmental or core temperature can impair the ANS, with all due complications (Fox and Matthews, 1989; Mowery et al., 2011).

Nonetheless, when used correctly, the temperature can also help newborns to correctly develop: indeed, therapeutic hypothermia in newborns with 36 WGA in case of HIE can help reduce brain and injury brain inflammation, thus having positive effects on neurodevelopment (Massaro et al., 2017; Lemyre and Chau, 2018).

### 3.3.2. Noise

If born prematurely, newborns lose the low-frequency maternal sounds that are essential for the correct maturation of their hearing system, with negative consequences on brain maturation and speech development (McMahon et al., 2012).

Furthemore, without uterus protection, newborns can be particularly sensitive to sounds, which usually overcome the 45dB limit recommended by the American Academy of Pediatrics (Almadhoob and Ohlsson, 2020), thus resulting in toxic stress (Weber and Harrison, 2019). The consequences can be quite severe: from apnoea to hypoxemia, from sleep disruption to decreased growth, from immunity impairment to neuroendocrine alterations (Almadhoob and Ohlsson, 2020).

Therefore, every possible procedure aimed to protect babies from harmful noises or to introduce positive sounds including the human voice, especially maternal voice (even recorded), and songs, can result in important neuroprotective effects (McMahon et al., 2012; Weber and Harrison, 2019). It is noteworthy that earmuffs show ambiguous results, that is, they can both help oxygenation and sleep and provoke stress (Zahr and de Traversay, 1995; Aita et al., 2013; Almadhoob and Ohlsson, 2020).

### 3.3.3. Light

As newborns pass the majority of their time sleeping, excessive light due to incubators not properly covered can negatively affect the newborns' cardiorespiratory regulation, thus increase stress level and altering ANS development (Ozawa et al., 2010; Weber and Harrison, 2019), in addition, to induce all the negative effects of sleep deficiency. Moreover, excessive light can impair visual development, thus increasing SNS and HPA axis activation (Weber and Harrison, 2019). On the other side, properly covering incubators or using eye masks facilitate a quiet sleep state characterized by a stable respiratory function (Shiroiwa et al., 1986; Venkataraman et al., 2018).

### 3.3.4. Electromagnetic fields

Newborns can also be subject to EMFs, from the incubator or the care unit environment. EMF coming from the incubator power influences the ANS as revealed by HRV analysis as LF/HF increases, whereas HF decreases (Bellieni et al., 2008). The farther away newborns stay from the incubator power, the less the ANS is affected (Bellieni et al., 2008; Passi et al., 2017).

Although incubator EMF seems to be under the normative value, EMFs could become dangerous when the incubator, and thus the newborn, is surrounded by other instruments or staff's devices that, through their own EMF, can give rise to electromagnetic interference (Besset et al., 2020). This phenomenon has the potential of altering ANS in three ways, especially in preterm or vulnerable infants. Firstly, by directly affecting brain development, as the skull has a low bone density that fails in adequately protecting newborns' brain from EMF (Besset et al., 2020). Secondly, by suppressing melatonin secretion, thus altering the immune system and, potentially, the CAP, with all the negative consequences already mentioned. Lastly, instrumental malfunctioning can induce life-threatening conditions (Carvajal de la Osa et al., 2020).

### 3.3.5. Incorrect oxygenation

Correct oxygenation is paramount for the newborn's development and survival (Ho et al., 2020). Indeed, in preterm infants, both too high or too low oxygenation can be dangerous for the nervous system. For instance, too high oxygenation can easily induce lung injury, thus increasing inflammation and stressing the already immature CAP, whereas too low oxygenation can cause neurodevelopmental impairment and HIE (Rantakari et al., 2021).

It is noteworthy that delayed umbilical cord clamping, that is, delaying the procedure by 60-180s seems to improve both peripheral and cerebral oxygenation, in addition to several hemodynamics parameters, which could reveal a positive effect on ANS function and development (Bruckner et al., 2021).

### 3.3.6. Posture

Babies' position is another factor that could impact ANS.

Indeed, a prone position improves oxygenation, increases PNS tone, and lowers both salivary cortisol and stress behavior (Gomes et al., 2019). Nevertheless, prone sleeping is considered a risk factor for SIDS and it shows a lower short-term variability in HRV (Lucchini et al., 2016). It is thus noteworthy that, in the case of prone sleeping position, between the 2nd and 4th months after birth, both blood pressure and cerebral oxygenation seem to decrease, whereas there is a propensity for cortical arousal, maybe to counter the fall in pressure and oxygenation. This would explain why ANS impairment, for instance, due to prematurity, can make newborns more vulnerable to SIDS (Horne 2018). As supporting evidence, complications including fetal growth restriction alter sleep stages by reducing active sleep (Aldrete-Cortez et al., 2019).

Supine position with manual restraint for flexion seems to have positive effects on ANS as measured through HRV (Gomes et al., 2019). In the same way, side position during feeding seems to have stabilizing effects on both ANS, as revealed by HR, and oxygenation in preterm newborns (Thoyre et al., 2012), although results are conflicting (Raczyńska and Gulczyńska, 2019).

### 3.3.7. Respect for the baby's sleep

Sleep impairment due to prematurity or environmental factors can lead to global neurodevelopmental complications (e.g., attention deficit, motor disability, and low cognitive abilities), as newborns show weaker functional connectivity networks during such impaired sleep (Uchitel et al., 2021).

Drugs including theophylline, caffeine, and methylxanthine, used to support respiratory functions have the potential of altering sleep quality and induce drowsiness. In the same way, improper light and noise can disrupt newborns' sleep duration, especially in preterm and very preterm ones (Gogou et al., 2019). But maybe, more than anything, medical procedures tend to impair babies' sleep the most, in particular since in preterm newborns many interventions are perceived as noxious and painful, thus heightening the stress response (Holsti et al., 2006; Weber and Harrison, 2019). Luckily, applying gentle touch, an "intervention" that stimulates the vagus nerve via C-tactile fibers, to medical care has the potential to regulate ANS function (Manzotti et al., 2019) and, thus, it is conceivable it could have protective effects on sleep. Otherwise, even clustered-care could help reduce the time the babies' sleep is disrupted (Holsti et al., 2006).

# 4. Clinical implications for the management of fetuses and newborns

*"The newborn is not new, and neither is the fetus".*
*Paraphrased from a personal communication with Dr. Barry Schifrin*

Our current clinical management of fetal and maternal health is not individualized and intermittent, suffering from a pattern of catching up with clinical symptoms instead of using biometric data to predict and steer developmental trajectories and health outcomes for the pregnant mother and her fetus during the delivery and after to their fullest potential. The fear of missing fetal injury during labor has led to soaring cesarean section rates and the failure of the EFM has resulted in the misguided dissent over its utility in general throwing the proverbial baby with the bathwater.

What is amiss is the understanding and acting upon the fact that each fetus and newborn are not truly new, but, rather, come with weeks and months of experience in utero. Can we learn to assess this individual experience?

In this review, we highlighted the growing understanding of the vagus nerve's developmental physiology, the largest and perhaps most important homeostatic (or, better, homeokinetic) regulatory system in our body (Macklem and Seely, 2010; Herry et al., 2019). The exciting take-home message is that much of this development can be gauged non-invasively during pregnancy using readily available bioelectric sensors such as ECG or, less qualitatively well, but still useful, using ultrasound-derived heart rate estimation, the mainstay of today's CTG and EFM technologies that is in sore need of technological disruption.

What could and should we do to disrupt this status quo and enable individualized insights into the various developmental effects the vagus nerve maturation and pleiotropic physiology provide?

Research is increasingly pointing towards the existence of an "HRV code," that is, specific HRV spatio-temporal alterations related to specific organic alterations (e.g., hypoxia, cerebral, cardiac, or gut pathologies) (Herry et al., 2019; Frasch, 2020), to the impact of external factors (e.g., smoking, physical activity, diabetes) and to the fetal brain and nervous development (Hoyer et al., 2012, 2013a, 2015, 2017, 2019). Indeed, fHRV can be affected by both neural factors (i.e., ANS) and non-neural factors, e.g., metabolic and organic functions (DiPietro et al., 2015).

Therefore, knowing that the ANS-organ connections are established so early in the embryo could offer useful insight for comprehending the complex role of the vagus nerve and autonomic ganglia in the developing embryo/fetus. Currently, ANS development is usually assessed during the 3rd trimester, when vagus nerve myelination begins (Mulkey and du Plessis, 2018). However, the literature shows that the ANS can be assessed through fHRV from 15-20 WGA (Hoyer et al., 2012, 2013b, 2015, 2019; Shuffrey et al., 2019). Moreover, the heartbeat can be heard through ultrasound as early as the 6th WGA, the heart rate accelerates constantly during the first 8 WGA (Rodgers et al., 2015) and it was recently discovered that an intrinsic fHRV (i.e., due to sino-atrial node rhythms) exists – it can affect the global fHRV and be influenced by pregnancy events (e.g., hypoxia) (Frasch et al., 2020a) and it shows a specific "signature", that is, specific HRV metrics (Herry et al., 2019). Maternal and transabdominal ECG recorded in the early third trimester carry information about the chronic exposure of the mother-fetus dyad to stress detectable by deep learning, a form of artificial intelligence techniques (Sarkar et al., 2021). Therefore, it would be worth it to improve fetal cardiac monitoring technology in order to detect ECG and fHRV even in the first weeks of pregnancy.

Such an advancement could improve the embryo's health assessment during early pregnancy, thus going well beyond detecting embryonic-fetal malformations and toward individualized prospective functional assessments. For instance, although we know that folate deficiency can impede the neural tube closure or that alcohol can induce dramatic brain modifications (Moore et al., 2016), we still lack knowledge about more subtle influences, that is, the precise effects of just some alcohol during the first weeks of pregnancy or the early effects of maternal stress (Avalos et al., 2014; Caspers Conway et al., 2014; Antonelli et al., 2021). On the other hand, we are learning that certain factors (e.g., choline supplementation) may protect the fetus from excessive HPA axis activation and have long-lasting (into adulthood) effects on cognition through neural and epigenetic modifications (Korsmo et al., 2019). Furthermore, some studies have already shown the feasibility of correlating external factors related to maternal lifestyle or "internal" factors such as fetal movement to fHRV changes as early as 20-28 WGA (DiPietro et al., 2007; Hoyer et al., 2019; Shuffrey et al., 2019).

As these findings are available for the 2nd and 3rd trimester, assessing fHRV in the 1st trimester may shed more light on what can protect or damage the ANS in these early formative stages of development and the consequences of its alterations.

Taken together, non-invasive assessment of maternal and fetal health via HRV monitoring throughout the pregnancy, from its earliest stages, will enable an individualized risk profiling and outcome projection during the latter stages of gestation and during delivery, a true disruption of the status quo that will improve maternal and perinatal health outcomes.

## 5. Conflict of Interest

M. G. Frasch has patents on aECG (WO2018160890A1) and EEG technologies for fetal monitoring (US9215999).
The authors declare that the research was conducted in the absence of any other commercial or financial relationships that could be construed as a potential conflict of interest.

## 6. Author contributions

FC, MC, SV, CV, AM conceptualised the paper. FC, MC, SV, CV, MCA and MGF wrote the first draft. All authors reviewed, edited and approved the final version of the paper.

## 7. Funding

This research did not receive any specific grant from funding agencies in the public, commercial, or not-for-profit sectors. MCA is recipient of a Hans Fischer Senior Fellowship from IAS-TUM (Institute for Advanced Study-TUM, Munich).

## 8. Abbreviations

ABP: arterial blood pressure
ANS: autonomic nervous system
ASD: autism spectrum disorders
CAM: cardiac autonomic modulation
CAN: central autonomic network
CAP: cholinergic anti-inflammatory pathway
CNS: central nervous system
CRH: corticotropin-releasing hormone
CS: cesarean section
CTG: cardiotocography
ECG: electrocardiography
EEG: electroencephalography
EFM: electronic fetal monitoring
EMF: electromagnetic field
ENS: enteric nervous system
FASD: fetal alcohol spectrum disorders
fHR: fetal heart rate
fHRV: fetal heart rate variability
FM: fetal movement
GM: general movement
HIE: hypoxic-ischemic encephalopathy
HPA: hypothalamic-pituitary-adrenal
HR: heart rate
HRV: heart rate variability
NAS: neonatal abstinence syndrome
NC: neural crest
NEC: necrotizing enterocolitis
NOWS: neonatal opioid withdrawal syndrome
PAE: prenatal alcohol exposure
PGP: pelvic girdle pain
PNE: prenatal nicotine exposure
PNS: parasympathetic nervous system
REM: rapid eye movements
SDB: sleep-disordered breathing
sGC: synthetic glucocorticoid
SIDS: sudden infant death syndrome
SNS: sympathetic nervous system
sOT: synthetic oxytocin
SSRI: selective serotonin reuptake inhibitor
VAS: vibroacoustic stimulation

WGA: week gestational age
ZIKV: Zika virus

Cibils, L. A., and Votta, R. (1993). Clinical significance of fetal heart rate patterns during labor. IX: Prolonged pregnancy. *J. Perinat. Med.* 21, 107–116. doi:10.1515/jpme.1993.21.2.107.

Clarkson, C. E., and Adams, N. (2018). A qualitative exploration of the views and experiences of women with Pregnancy related Pelvic Girdle Pain. *Physiotherapy* 104, 338–346. doi:10.1016/j.physio.2018.05.001.

Cogan, R., and Spinnato, J. A. (1986). Pain and discomfort thresholds in late pregnancy. *Pain* 27, 63–68. doi:10.1016/0304-3959(86)90223-X.

Committee on fetus and newborn and section on anesthesiology and pain medicine (2016). Prevention and Management of Procedural Pain in the Neonate: An Update. *Pediatrics* 137, e20154271–e20154271. doi:10.1542/peds.2015-4271.

Conradt, E., Crowell, S. E., and Lester, B. M. (2018). Early life stress and environmental influences on the neurodevelopment of children with prenatal opioid exposure. *Neurobiol. Stress* 9, 48–54. doi:10.1016/j.ynstr.2018.08.005.

Conradt, E., Flannery, T., Aschner, J. L., Annett, R. D., Croen, L. A., Duarte, C. S., et al. (2019). Prenatal Opioid Exposure: Neurodevelopmental Consequences and Future Research Priorities. *Pediatrics* 144. doi:10.1542/peds.2019-0128.

Conway, C. R., Sheline, Y. I., Chibnall, J. T., George, M. S., Fletcher, J. W., and Mintun, M. A. (2006). Cerebral blood flow changes during vagus nerve stimulation for depression. *Psychiatry Res. Neuroimaging* 146, 179–184. doi:10.1016/j.pscychresns.2005.12.007.

Corsello, S. M., and Paragliola, R. M. (2019). Evaluation and Management of Endocrine Hypertension During Pregnancy. *Endocrinol. Metab. Clin. North Am.* 48, 829–842. doi:10.1016/j.ecl.2019.08.011.

Crudo, A., Suderman, M., Moisiadis, V. G., Petropoulos, S., Kostaki, A., Hallett, M., et al. (2013). Glucocorticoid programming of the fetal male hippocampal epigenome. *Endocrinology* 154, 1168–1180. doi:10.1210/en.2012-1980.

Davis, D. C. (1996). The Discomforts of Pregnancy. *J. Obstet. Gynecol. Neonatal Nurs.* 25, 73–81. doi:10.1111/j.1552-6909.1996.tb02516.x.

De Jonckheere, J., and Storme, L. (2019). NIPE is related to parasympathetic activity. Is it also related to comfort? *J. Clin. Monit. Comput.* 33, 747–748. doi:10.1007/s10877-019-00276-1.

De las Cuevas, C., and Sanz, E. J. (2006). Safety of Selective Serotonin Reuptake Inhibitors in Pregnancy. *Curr. Drug Saf.* 1, 17–24. doi:10.2174/157488606775252593.

De Rogalski Landrot, I., Roche, F., Pichot, V., Teyssier, G., Gaspoz, J.-M., Barthelemy, J.-C., et al. (2007). Autonomic nervous system activity in premature and full-term infants from theoretical term to 7 years. *Auton. Neurosci.* 136, 105–109. doi:10.1016/j.autneu.2007.04.008.

de Vries, J. I. P., Visser, G. H. A., and Prechtl, H. F. R. (1982). The emergence of fetal behaviour. I. Qualitative aspects. *Early Hum. Dev.* 7, 301–322. doi:10.1016/0378-3782(82)90033-0.

de Vries, J. I. P., Visser, G. H. A., and Prechtl, H. F. R. (1985). The emergence of fetal behaviour. II. Quantitative aspects. *Early Hum. Dev.* 12, 99–120. doi:10.1016/0378-3782(85)90174-4.

Dejmek, J., Selevan, S. G., Benes, I., Solanský, I., and Srám, R. J. (1999). Fetal growth and maternal exposure to particulate matter during pregnancy. *Environ. Health Perspect.* 107, 475–480. doi:10.1289/ehp.99107475.

Derbyshire, S. W., and Bockmann, J. C. (2020). Reconsidering fetal pain. *J. Med. Ethics* 46, 3–6. doi:10.1136/medethics-2019-105701.

Dereymaeker, A., Pillay, K., Vervisch, J., De Vos, M., Van Huffel, S., Jansen, K., et al. (2017). Review of sleep-EEG in preterm and term neonates. *Early Hum. Dev.* 113, 87–103. doi:10.1016/j.earlhumdev.2017.07.003.

Desrosiers, T. A., Lawson, C. C., Meyer, R. E., Stewart, P. A., Waters, M. A., Correa, A., et al. (2015). Assessed occupational exposure to chlorinated, aromatic and Stoddard solvents during pregnancy and risk of fetal growth restriction. *Occup. Environ. Med.*
38

313. doi:10.3389/fphys.2017.00313.

Massaro, A. N., Campbell, H. E., Metzler, M., Al-Shargabi, T., Wang, Y., du Plessis, A., et al. (2017). Effect of temperature on heart rate variability in neonatal intensive care unit patients with hypoxic ischemic encephalopathy. *Pediatr. Crit. Care Med.* 18, 349–354. doi:10.1097/PCC.0000000000001094.

Mattingly, J. E., D'Alessio, J., and Ramanathan, J. (2003). Effects of Obstetric Analgesics and Anesthetics on the Neonate. *Pediatr. Drugs* 5, 615–627. doi:10.2165/00148581-200305090-00004.

McCarthy, J. J., Leamon, M. H., Finnegan, L. P., and Fassbender, C. (2017). Opioid dependence and pregnancy: minimizing stress on the fetal brain. *Am. J. Obstet. Gynecol.* 216, 226–231. doi:10.1016/j.ajog.2016.10.003.

McEwen, B. S., Nasca, C., and Gray, J. D. (2016). Stress Effects on Neuronal Structure: Hippocampus, Amygdala, and Prefrontal Cortex. *Neuropsychopharmacol. Off. Publ. Am. Coll. Neuropsychopharmacol.* 41, 3–23. doi:10.1038/npp.2015.171.

McKenna, D. S., Ventolini, G., Neiger, R., and Downing, C. (2006). Gender-related differences in fetal heart rate during first trimester. *Fetal Diagn. Ther.* 21, 144–147. doi:10.1159/000089065.

McMahon, E., Wintermark, P., and Lahav, A. (2012). Auditory brain development in premature infants: the importance of early experience: McMahon et al. *Ann. N. Y. Acad. Sci.* 1252, 17–24. doi:10.1111/j.1749-6632.2012.06445.x.

Mercer, B. M., Carr, T. L., Beazley, D. D., Crouse, D. T., and Sibai, B. M. (1999). Antibiotic use in pregnancy and drug-resistant infant sepsis. *Am. J. Obstet. Gynecol.* 181, 816–821. doi:10.1016/S0002-9378(99)70307-8.

Metzler, M., Govindan, R., Al-Shargabi, T., Vezina, G., Andescavage, N., Wang, Y., et al. (2017). Pattern of brain injury and depressed heart rate variability in newborns with hypoxic ischemic encephalopathy. *Pediatr. Res.* 82, 438–443. doi:10.1038/pr.2017.94.

Meyer, U., Yee, B. K., and Feldon, J. (2007). The Neurodevelopmental Impact of Prenatal Infections at Different Times of Pregnancy: The Earlier the Worse? *The Neuroscientist* 13, 241–256. doi:10.1177/1073858406296401.

Micheli, K., Komninos, I., Bagkeris, E., Roumeliotaki, T., Koutis, A., Kogevinas, M., et al. (2011). Sleep Patterns in Late Pregnancy and Risk of Preterm Birth and Fetal Growth Restriction. *Epidemiology* 22, 738–744.

Miljković, Z., Sabo, A., Stanulović, M., Jakovljević, V., and Grujić, I. (2001). [Drug use during pregnancy, labor and the puerperium and after the Drug Use in Pregnancy Study]. *Med. Pregl.* 54, 34–37.

Mindell, J. A., Cook, R. A., and Nikolovski, J. (2015). Sleep patterns and sleep disturbances across pregnancy. *Sleep Med.* 16, 483–488. doi:10.1016/j.sleep.2014.12.006.

Mirmiran, M., Maas, Y. G. H., and Ariagno, R. L. (2003). Development of fetal and neonatal sleep and circadian rhythms. *Sleep Med. Rev.* 7, 321–334. doi:10.1053/smrv.2002.0243.

Montagna, A., and Nosarti, C. (2016). Socio-Emotional Development Following Very Preterm Birth: Pathways to Psychopathology. *Front. Psychol.* 7, 80. doi:10.3389/fpsyg.2016.00080.

Moore, K. L., Persaud, T. V. N., and Torchia, M. G. (2016). *The developing human: clinically oriented embryology*. 10th edition. Philadelphia, PA: Elsevier.

Moors, S., Staaks, K. J. J., Westerhuis, M. E. M. H., Dekker, L. R. C., Verdurmen, K. M. J., Oei, S. G., et al. (2020). Heart rate variability in hypertensive pregnancy disorders: A systematic review. *Pregnancy Hypertens.* 20, 56–68. doi:10.1016/j.preghy.2020.03.003.

Moreira Neto, R., and Porovic, S. (2018). Clinical study of fetal neurobehavior by the KANET test. *J. Perinat. Med.* 46, 631–639. doi:10.1515/jpm-2016-0414.

Morris, G. S., Zhou, Q., Hegsted, M., and Keenan, M. J. (1995). Maternal consumption of a low vitamin D diet retards metabolic and contractile development in the neonatal rat heart. *J. Mol. Cell. Cardiol.* 27, 1245–1250. doi:10.1016/s0022-2828(05)82386-7.

doi:10.1111/j.1471-0528.2007.01236.x.

Wilder, R. L. (1998). Hormones, Pregnancy, and Autoimmune Diseases. *Ann. N. Y. Acad. Sci.* 840, 45–50. doi:https://doi.org/10.1111/j.1749-6632.1998.tb09547.x.

Wilson, J. D., Adams, A. J., Murphy, P., Eswaran, H., and Preissl, H. (2008). Design of a light stimulator for fetal and neonatal magnetoencephalography. *Physiol. Meas.* 30, N1–N10. doi:10.1088/0967-3334/30/1/N01.

Woodson, R. H., Jones, N. G., da Costa Woodson, E., Pollock, S., and Evans, M. (1979). Fetal mediators of the relationships between increased pregnancy and labour blood pressure and newborn irritability. *Early Hum. Dev.* 3, 127–139. doi:10.1016/0378-3782(79)90002-1.

Yee, J. R., Kenkel, W. M., Frijling, J. L., Dodhia, S., Onishi, K. G., Tovar, S., et al. (2016). Oxytocin promotes functional coupling between paraventricular nucleus and both sympathetic and parasympathetic cardioregulatory nuclei. *Horm. Behav.* 80, 82–91. doi:10.1016/j.yhbeh.2016.01.010.

Yiallourou, S. R., Sands, S. A., Walker, A. M., and Horne, R. S. C. (2012). Maturation of Heart Rate and Blood Pressure Variability during Sleep in Term-Born Infants. *SLEEP*. doi:10.5665/sleep.1616.

Yurdakök, K. (2012). View of Environmental pollution and the fetus. *J. Pediatr. Neonatal Individ. Med. JPNIM* 1, 33–42. doi:10.7363/010116.

Zafarghandi, N., Hadavand, S., Davati, A., Mohseni, S. M., Kimiaiimoghadam, F., and Torkestani, F. (2012). The effects of sleep quality and duration in late pregnancy on labor and fetal outcome. *J. Matern. Fetal Neonatal Med.* 25, 535–537. doi:10.3109/14767058.2011.600370.

Zahr, L. K., and de Traversay, J. (1995). Premature infant responses to noise reduction by earmuffs: effects on behavioral and physiologic measures. *J. Perinatol. Off. J. Calif. Perinat. Assoc.* 15, 448–455.

Zeskind, P. S., and Gingras, J. L. (2006). Maternal Cigarette-Smoking During Pregnancy Disrupts Rhythms in Fetal Heart Rate. *J. Pediatr. Psychol.* 31, 5–14. doi:10.1093/jpepsy/jsj031.

Zeskind, P. S., and Stephens, L. E. (2004). Maternal Selective Serotonin Reuptake Inhibitor Use During Pregnancy and Newborn Neurobehavior. *Pediatrics* 113, 368–375. doi:10.1542/peds.113.2.368.

Zhang, J., Cai, W.-W., and Lee, D. J. (1992). Occupational hazards and pregnancy outcomes. *Am. J. Ind. Med.* 21, 397–408. doi:https://doi.org/10.1002/ajim.4700210312.

Zhang, X., Sliwowska, J. H., and Weinberg, J. (2005). Prenatal Alcohol Exposure and Fetal Programming: Effects on Neuroendocrine and Immune Function. *Exp. Biol. Med.* 230, 376–388. doi:10.1177/15353702-0323006-05.

Zheng, J., Xiao, X., Zhang, Q., Mao, L., Yu, M., and Xu, J. (2015). The Placental Microbiome Varies in Association with Low Birth Weight in Full-Term Neonates. *Nutrients* 7, 6924–6937. doi:10.3390/nu7085315.

Zhou, L., and Xiao, X. (2018). The role of gut microbiota in the effects of maternal obesity during pregnancy on offspring metabolism. *Biosci. Rep.* 38. doi:10.1042/BSR20171234.

Zizzo, A. R., Kirkegaard, I., Hansen, J., Uldbjerg, N., and Mølgaard, H. (2020). Fetal Heart Rate Variability Is Affected by Fetal Movements: A Systematic Review. *Front. Physiol.* 11, 578898. doi:10.3389/fphys.2020.578898.

**Table 1. Summary of the HRV metrics cited in the main text (Brändle et al., 2015; Massaro et al., 2017; Frasch et al., 2020; Oliveira et al., 2019; Patural et al., 2019).**

| Parameter | Definition |
|---|---|
| **Time-domain** | |
| SDNN | Standard deviation of NN intervals. Related to both SNS and PNS activity and equivalent to Poincaré SD2. |
| RMSSD | Root mean square of consecutive RR interval differences. Related to vagal modulation and equivalent to Poincaré SD1. |
| **Frequency-domain** | |
| LF | Low-frequency band: 0.04-0.2Hz for newborns and 0.04-0.15Hz for infants. It can be expressed as LF peak, LF power or LFn (normalized power in relation to total power). Related to both sympathetic and vagal modulation and baroreceptor reflexes. |
| HF | High-frequency band: 0.20-2.00Hz for newborns and 0.20-1.40Hz for infants. It can be expressed as HF peak, HF power or HFn (normalized power in relation to total power). Related to both vagal modulation and respiratory cycle. |
| LF/HF | Ratio of LF-to-HF power. Related to both SNS and PNS activity. |
| TP | Total power of the electrocardiography spectrogram. |
| **Nonlinear** | |
| SD1 | Poincaré plot standard deviation perpendicular to the line of identity. Equivalent to RMSSD. |
| SD2 | Poincaré plot standard deviation along the line of identity. Equivalent to SDNN. |
| CSI | Cardiac Sympathetic Index (ratio of SD2-to-SD1). |
| HRV Index | Number of all RR intervals divided by the number of RR intervals at the highest point of the RR histogram. |
| Parseval Index | Ratio between the square root of the sum of LF and HF powers and the value of SDNN |
| PE | Permutation entropy. |
| DFA $\alpha_1$ | Detrended fluctuation analysis, which describes short-term fluctuations. Usually used to assess brain injury or pathology severity. |
| DFA $\alpha_2$ | Detrended fluctuation analysis, which describes long-term fluctuations. Usually used to assess brain injury or pathology severity. |
| RMS1 | Root mean square from detrended fluctuation analysis, which describes short-term fluctuations. Usually used to assess brain injury or pathology severity. |
| RMS2 | Root mean square from detrended fluctuation analysis, which describes long-term fluctuations. Usually used to assess brain injury or pathology severity. |